\documentclass[preprint]{aastex}

\usepackage{amssymb}
\usepackage{amsmath}

\newcommand{\Msun}{M_\odot}
\newcommand{\Mmin}{M_{\textnormal{min}}}
\newcommand{\Mmax}{M_{\textnormal{max}}}
\newcommand{\Nbin}{N_{\textnormal{bin}}}
\newcommand{\vtheta}{\vec{\theta}}
\newcommand{\order}[1]{\mathcal{O}\left( #1 \right)}

\begin{document}

\title{The Mass Distribution of Stellar-Mass Black Holes}

\author{Will M. Farr \and Niharika Sravan} 

\affil{Northwestern University Center for Interdisciplinary
  Exploration and Research in Astrophysics\\2145 Sheridan Rd.,
  Evanston, IL 60208}

\email{w-farr@northwestern.edu, niharika.sravan@gmail.com}

\author{Andrew Cantrell \and Laura Kreidberg \and Charles D. Bailyn}

\affil{Yale University Department of Astrophysics\\
  P.O. Box 208101, New Haven, CT 06520}

\email{andrew.cantrell@yale.edu, laura.kreidberg@yale.edu,
  charles.bailyn@yale.edu}

\author{Ilya Mandel\altaffilmark{1}}

\affil{MIT Kavli Institute, Cambridge, MA 02139}

\email{ilyamandel@chgk.info}

\and

\author{Vicky Kalogera} 

\affil{Northwestern University Center for Interdisciplinary
  Exploration and Research in Astrophysics\\2145 Sheridan Rd.,
  Evanston, IL 60208}

 \email{vicky@northwestern.edu}

 \altaffiltext{1}{Also: School of Physics and Astronomy, University of
   Birmingham, Edgbaston, Birmingham B15 2TT, UK}

\begin{abstract}
  We perform a Bayesian analysis of the mass distribution of
  stellar-mass black holes using the observed masses of 15 low-mass
  X-ray binary systems undergoing Roche lobe overflow and five
  high-mass, wind-fed X-ray binary systems.  Using Markov Chain
  Monte Carlo calculations, we model the mass distribution both
  parametrically---as a power law, exponential, gaussian, combination
  of two gaussians, or log-normal distribution---and
  non-parametrically---as histograms with varying numbers of bins.  We
  provide confidence bounds on the shape of the mass distribution in
  the context of each model and compare the models with each other by
  calculating their relative Bayesian evidence as supported by the
  measurements, taking into account the number of degrees of freedom
  of each model.  The mass distribution of the low-mass systems is
  best fit by a power-law, while the distribution of the combined
  sample is best fit by the exponential model.  This difference
  indicates that the low-mass subsample is not consistent with being
  drawn from the distribution of the combined population.  We examine
  the existence of a ``gap'' between the most massive neutron stars
  and the least massive black holes by considering the value,
  $M_{1\%}$, of the 1\% quantile from each black hole mass
  distribution as the lower bound of black hole masses.  Our analysis
  generates posterior distributions for $M_{1\%}$; the best model (the
  power law) fitted to the low-mass systems has a distribution of
  lower-bounds with $M_{1\%} > 4.3$ $\Msun$ with 90\% confidence,
  while the best model (the exponential) fitted to all 20 systems has
  $M_{1\%} > 4.5$ $\Msun$ with 90\% confidence.  We conclude that our
  sample of black hole masses provides strong evidence of a gap
  between the maximum neutron star mass and the lower bound on black
  hole masses.  Our results on the low-mass sample are in qualitative
  agreement with those of \citet{Ozel2010}, although our broad
  model-selection analysis more reliably reveals the best-fit
  quantitative description of the underlying mass distribution.  The
  results on the combined sample of low- and high-mass systems are
  in qualitative agreement with \citet{Fryer2001} although the
  presence of a mass gap remains theoretically unexplained.
\end{abstract}


\maketitle

\section{Introduction}
\label{sec:intro}

The most massive stars probably end their lives with a supernova
explosion or a quiet core collapse, becoming stellar-mass black holes.
The mass distribution of such black holes can provide important clues
to the end stages of evolution of these stars.  In addition, the mass
distribution of stellar-mass black holes is an important input in
calculations of rates of gravitational wave emission events from
coalescing neutron star-black hole and black hole-black hole binaries
in the LIGO gravitational wave observatory \citep{Abadie2010}.

Observations of X-ray binaries in both the optical and X-ray bands can
provide a measurement of the mass of the compact object in these
systems.  The current sample of stellar mass black holes with
dynamically measured masses includes 15 systems with low-mass, Roche
lobe overflowing donors and 5 wind-fed systems with high-mass
donors.  Hence, sophisticated statistical analyses of the black hole
mass distribution in these systems are possible.

The first study of the mass distribution of stellar-mass black holes,
in \citet{Bailyn1998}, examined a sample of seven low-mass X-ray
binaries thought to contain a black hole, concluding in a Bayesian
analysis that the mass function was strongly peaked around seven solar
masses%
\footnote{A similar analysis of the neutron star mass distribution can
  be found in \citet{Finn1994}.}. %
\citet{Bailyn1998} found evidence of a ``gap'' between the least
massive black hole and a ``safe'' upper limit for neutron star masses
of 3 $\Msun$ (e.g.\ \citet{Kalogera1996}).  Such a gap is puzzling in
light of theoretical studies that predict a continuous distribution of
compact object supernova remnant masses with a smooth transition from
neutron stars to black holes \citep{Fryer2001}.  (We note that
\citet{Fryer2001} considered binary evolution effects only
heuristically and put forward some possible explanations for the gap
from \citet{Bailyn1998} both in the context of selection effects or in
connection to the energetics of supernova explosions.)

Towards the end of our analysis work, we became aware of a more recent
study \citep{Ozel2010}, also in a Bayesian framework, analyzing the
low-mass X-ray binary sample.  Our results are largely consistent with
those obtained by \citet{Ozel2010}, who examined 16 low-mass X-ray
binary systems containing black holes and found a strongly peaked
distribution at $7.8 \pm 1.2 \, \Msun$.  They used two models for the
mass function: a Gaussian and a decaying exponential with a minimum
``turn-on'' mass (motivated by the analytical model of the black-hole
mass function in \citet{Fryer2001}).  We note that \citet{Ozel2010} do
not provide confidence limits for the minimum black hole mass, instead
discussing only the model parameters at the peak of their posterior
distributions.  They also do not perform any model selection analysis;
thus, they give the distribution of parameters within each of their
models, but cannot say which model is more likely to correspond to the
true distribution of black hole masses.  Nevertheless, it appears that
their analysis confirms the existence of a mass gap.  \citet{Ozel2010}
discuss possible selection effects that could lead to the appearance
of a mass gap, but conclude these effects could not produce the
observed gap, which they therefore claim is a real property of the
black hole mass distribution.

We use a Bayesian Markov-Chain Monte Carlo (MCMC) analysis to
quantitatively assess a wide range of models for the black hole mass
function for both samples.  We include both parametric models, such
as a Gaussian, and non-parametric models where the mass function is
represented by histograms with various numbers of bins.  (Our set of
models includes those of \citet{Ozel2010} and \citet{Bailyn1998}.)
After computing posterior distributions for the model parameters, we
use model selection techniques (including a new technique for
efficient reversible-jump MCMC \citep{Farr2010}) to compare the
evidence for the various models from both samples.

We define the ``minimum black hole mass'' to be the 1\% quantile,
$M_{1\%}$, in the black hole mass distribution (see Section
\ref{sec:minimum-mass}).  In qualitative agreement with
\citet{Ozel2010} and \citet{Bailyn1998}, we find strong evidence for a
mass gap among the best models for both samples.  Our analysis gives
distributions for $M_{1\%}$ implied by the data in the context of each
of our models for the black hole mass distribution.  In the context of
the best model for the low-mass systems (a power-law), the
distribution for $M_{1\%}$ gives $M_{1\%} > 4.3$ $\Msun$ with 90\%
confidence; in the context of the best model for the combined sample
of lower- and high-mass systems the distribution of $M_{1\%}$ has
$M_{1\%} > 4.5$ $\Msun$ with 90\% confidence.  Further, in the context
of models with lower evidence, most also have a mass gap, with 90\%
confidence bounds on $M_{1\%}$ significantly above a ``safe'' maximum
neutron star mass of 3 $\Msun$ \citep{Kalogera1996}.

We find that, for the low-mass X-ray binary sample, the theoretical
model from \citet{Fryer2001}---a decaying exponential---is strongly
disfavored by our model selection.  We find that the low-mass systems
are best described by a power law, followed closely by a Gaussian
(which is the second model considered by \citet{Ozel2010}).  On the
other hand, we find that the theoretical model from \citet{Fryer2001}
is the preferred model for the combined sample of low- and high-mass
X-ray binaries.  A model with two separate Gaussian peaks also has
relatively high evidence for the combined sample of systems.  The
difference in best-fitting model indicates that the low-mass subsample
is not consistent with being drawn from the distribution of the
combined population.

The structure of this paper is as follows.  In Section
\ref{sec:systems} we discuss the 15 systems that comprise the low-mass
X-ray binary black hole sample and the 5 additional high-mass,
wind-fed systems that make up the combined sample.  In Section
\ref{sec:models} we discuss the Bayesian techniques we use to analyze
the black hole mass distribution, the techniques we use for model
selection, and the parametric and non-parametric models we will use
for the black hole mass distribution.  In Section \ref{sec:results} we
discuss the results of our analysis and model selection.  In Section
\ref{sec:minimum-mass} we discuss the distribution of the minimum
black hole mass implied by the analysis of Section \ref{sec:results}.
In Section \ref{sec:conclusion} we summarize our results and comment
on the significance of the observed mass gap in the context of
theoretical models.  Appendix \ref{sec:mcmc} describes MCMC techniques
in some detail.  Appendix \ref{sec:reversible-jump-mcmc} explains our
novel algorithm for efficiently performing the reversible jump MCMC
computations used in the model comparisons of Section
\ref{sec:results} (but see also \citet{Farr2010}).

\section{Systems}
\label{sec:systems}

The 20 X-ray binary systems on which this study is based are listed in
Table \ref{tab:sources}.  We separate the systems into 15 low-mass
systems in which the central black hole appears to be fed by
roche-lobe overflow from the secondary, and 5 high-mass systems in
which the black hole is fed via winds (these systems all have a
secondary that appears to be more massive than the black hole).  The
low- and high-mass systems undoubtedly have different evolutionary
tracks, and therefore it is reasonable that they would have different
black-hole mass distributions.  We will first analyze the 15 low-mass
systems alone (Section \ref{sec:results-low-mass}), and then the
combined sample of 20 systems (Section \ref{sec:high-mass}).

In each of these systems, spectroscopic measurements of the secondary
star provide an orbital period for the system and a semi-amplitude for
the secondary's velocity curve.  These measurements can be combined
into the mass function,
\begin{equation}
  \label{eq:mass-function}
  f(M) = \frac{P K^3}{2\pi G} = \frac{M \sin^3 i}{\left( 1 + q \right)^2},
\end{equation}
where $P$ is the orbital period, $K$ is the secondary's velocity
semi-amplitude, $M$ is the black hole mass, $i$ is the inclination of
the system, and $q \equiv M_2 / M$ is the mass ratio of the system.

The mass function defines a lower limit on the mass: $f(M) < M$.  To
accurately determine the mass of the black hole, the inclination $i$
and mass ratio $q$ must be measured.  Ideally, this can be
accomplished by fitting ellipsoidal light curves and study of the
rotational broadening of spectral lines from the secondary, but even
in the most studied case (see, e.g., \citet{Cantrell2010} on A0620)
this procedure is complicated.  In particular, contributions from an
accretion disk and hot spots in the disk can significantly distort the
measured inclination and mass ratios.  For some systems (e.g.\ GS 1354
\citep{Casares2009}) strong variability completely prevents
determination of the inclination from the lightcurve; in these cases
an upper limit on the inclination often comes from the observed lack
of eclipses in the lightcurve.  In general, accurately determining $q$
and $i$ requires a careful system-by-system analysis.

For the purposes of this paper, we adopt the following simplified
approach to the estimation of the black hole mass from the observed
data.  When an observable is well-constrained, we assume that the true
value is normally distributed about the measured value with a standard
deviation equal to the quoted observational error.  This is the case
for the mass function in all the systems we use, and for many systems'
mass ratios and inclinations.  When a large range is quoted in the
literature for an observable, we take the true value to be distributed
uniformly (for the mass ratio) or isotropically (for the inclination)
within the quoted range.  Table \ref{tab:sources} gives the assumed
distribution for the observables in the 20 systems we use.  We do not
attempt to deal with the systematic biases in the observational
determination of $f$, $q$, and $i$ in any realistic way; we are
currently investigating more realistic treatments of the errors
(including observational biases that can shift the peak of the true
mass distribution away from the ``best-fit'' mass in the
observations).  This treatment will appear in future work.

\begin{table}
  \begin{center}
    \begin{tabular}{|l|c|c|c|l|}
      \hline
      Source & $f$ ($\Msun$) & $q$ & $i$ (degrees) & References \\
      \hline \hline
      GRS 1915 & $N(9.5, 3.0)$ & $N(0.0857, 0.0284)$ & $N(70, 2)$ &
      \citet{Greiner2001} \\
      XTE J1118 & $N(6.44, 0.08)$ & $N(0.0264, 0.004)$ & $N(68, 2)$ &
      \citet{Gelino2008} \\ & & & & \citet{Harlaftis2005} \\
      XTE J1650 & $N(2.73, 0.56)$ & $U(0, 0.5)$ & $I(50, 80)$ &
      \cite{Orosz2004} \\
      GRS 1009 & $N(3.17, 0.12)$ & $N(0.137, 0.015)$ & $I(37, 80)$ &
      \cite{Filippenko1999} \\
      A0620 & $N(2.76, 0.036)$ & $N(0.06, 0.004)$ & $N(50.98, 0.87)$ &
      \citet{Cantrell2010} \\ & & & & \citet{Neilsen2008} \\
      GRO J0422 & $N(1.13, 0.09)$ & $U(0.076, 0.31)$ & $N(45, 2)$ &
      \citet{Gelino2003} \\
      Nova Mus 1991 & $N(3.01, 0.15)$ & $N(0.128, 0.04)$ & $N(54,1.5)$
      & \cite{Gelino2001} \\
      GRO J1655 & $N(2.73,0.09)$ & $N(0.3663, 0.04025)$ & $N(70.2,
      1.9)$ & \citet{Greene2001} \\
      4U 1543 & $N(0.25, 0.01)$ & $U(0.25, 0.31)$ & $N(20.7,1.5)$ & 
      \citet{Orosz2003} \\
      XTE J1550 & $N(7.73,0.4)$ & $U(0,0.04)$ & $N(74.7, 3.8)$ &
      \citet{Orosz2010} \\
      V4641 Sgr & $N(3.13,0.13)$ & $U(0.42,0.45)$ & $N(75,2)$ &
      \citet{Orosz2003} \\
      GS 2023 & $N(6.08, 0.06)$ & $U(0.056,0.063)$ & $I(66, 70)$ &
      \citet{Charles2006} \\
      & & & & \citet{Khargharia2010} \\
      GS 1354 & $N(5.73, 0.29)$ & $N(0.12,0.04)$ & $I(50, 80)$ & 
      \citet{Casares2009} \\
      Nova Oph 77 & $N(4.86,0.13)$ & $U(0, 0.053)$ & $I(60, 80)$ &
      \citet{Charles2006} \\
      GS 2000 & $N(5.01, 0.12)$ & $U(0.035, 0.053)$ & $I(43, 74)$ &
      \citet{Charles2006} \\
      \hline \hline
      Cyg X1 & $N(0.251, 0.007)$ & $N(2.778, 0.386)$ & $I(23, 38)$ &
      \citet{Gies2003} \\ 
      M33 X7 & $N(0.46, 0.08)$ & $N(4.47, 0.61)$ & $N(74.6, 1)$ &
      \citet{Orosz2007} \\
      NGC 300 X1 & $N(2.6, 0.3)$ & $U(1.05, 1.65)$ & $I(60, 75)$ &
      \citet{Crowther2010} \\
      LMC X1 & $N(0.148, 0.004)$ & $N(2.91, 0.49)$ & $N(36.38, 2.02)$
      & \citet{Orosz2009} \\
      IC 10 X1 & $N(7.64, 1.26)$ & $U(0.7, 1.7)$ & $I(75, 90)$ & 
      \citet{Prestwich2007} \\
       & & & & \citet{Silverman2008} \\
       \hline
    \end{tabular}
  \end{center}
  \caption{\label{tab:sources} The source parameters for the 20 X-ray binaries used in this work.  The first 15 systems have
    low-mass secondaries that feed the black hole via Roche lobe
    overflow; the last five systems have high-mass secondaries ($q
    \gtrsim 1$) that feed the black hole via winds.  In each line, $f$
    is the mass function for
    the compact object, $q$ is the mass ratio $M_2/M$, and $i$ is the
    inclination of the system to the line of sight.  We indicate the
    distribution used for the true parameters when computing the
    probability distributions for the masses of these systems:
    $N(\mu,\sigma)$ implies a Gaussian with mean $\mu$ and standard
    deviation $\sigma$, $U(a,b)$ is a uniform distribution between $a$ and
    $b$, and $I(\alpha,\beta)$ is an isotropic distribution between the
    angles $\alpha$ and $\beta$.}
\end{table}

\begin{figure}
  \begin{center}
    \plotone{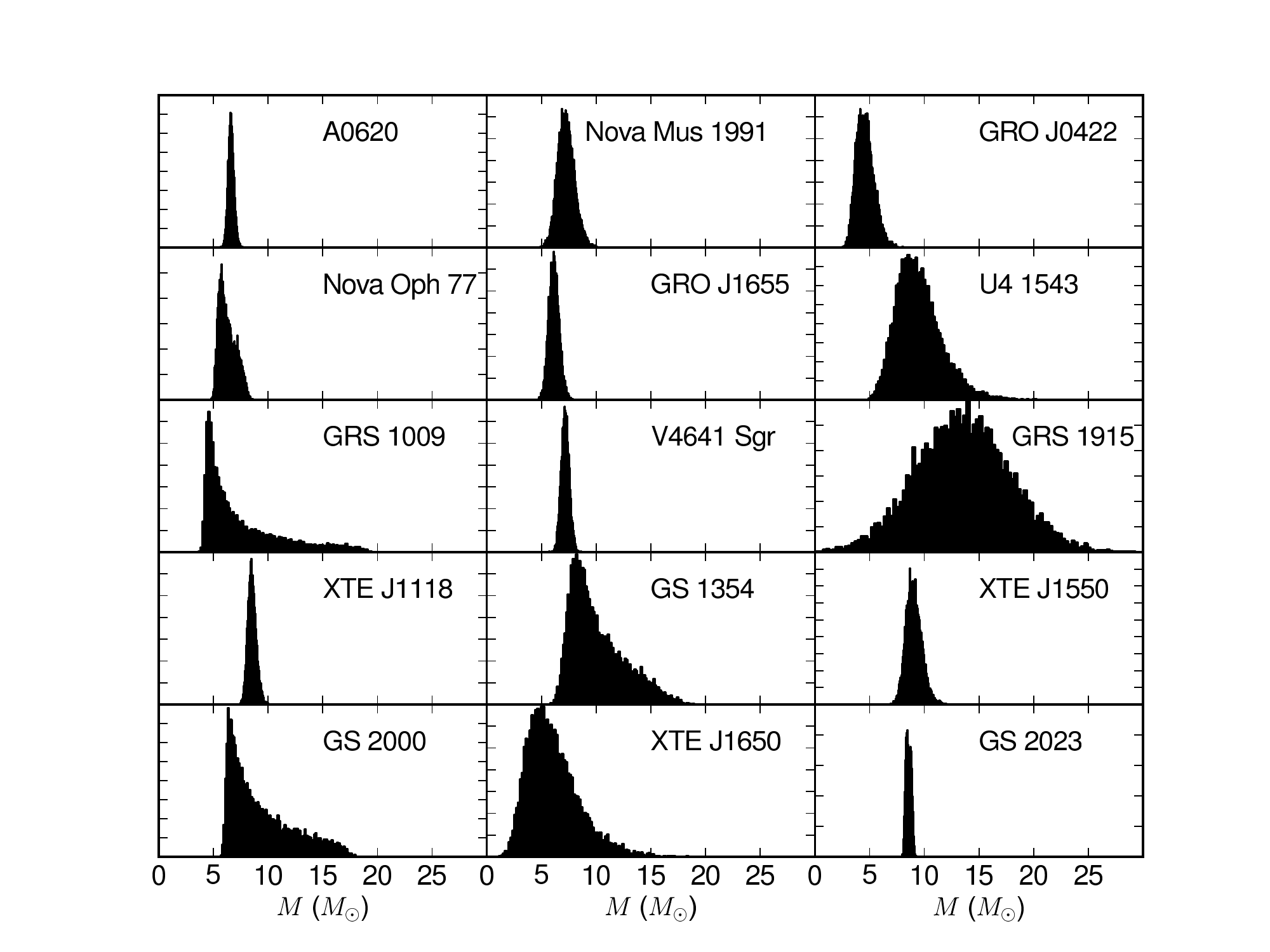} 
  \end{center}

  \caption{\label{fig:low-masses} The individual mass distributions
    implied by Equation \eqref{eq:mass-function} and the assumed
    distributions on observational parameters $f$, $q$, and $i$ given
    in Table \ref{tab:sources} for the low-mass sources.  The
    significant asymmetry and long tails in many of these
    distributions are the result of the non-linear relationship
    (Equation \eqref{eq:mass-function}) between $M$, $f$, $q$, and
    $i$.}
\end{figure}

\begin{figure}
  \begin{center}
    \plotone{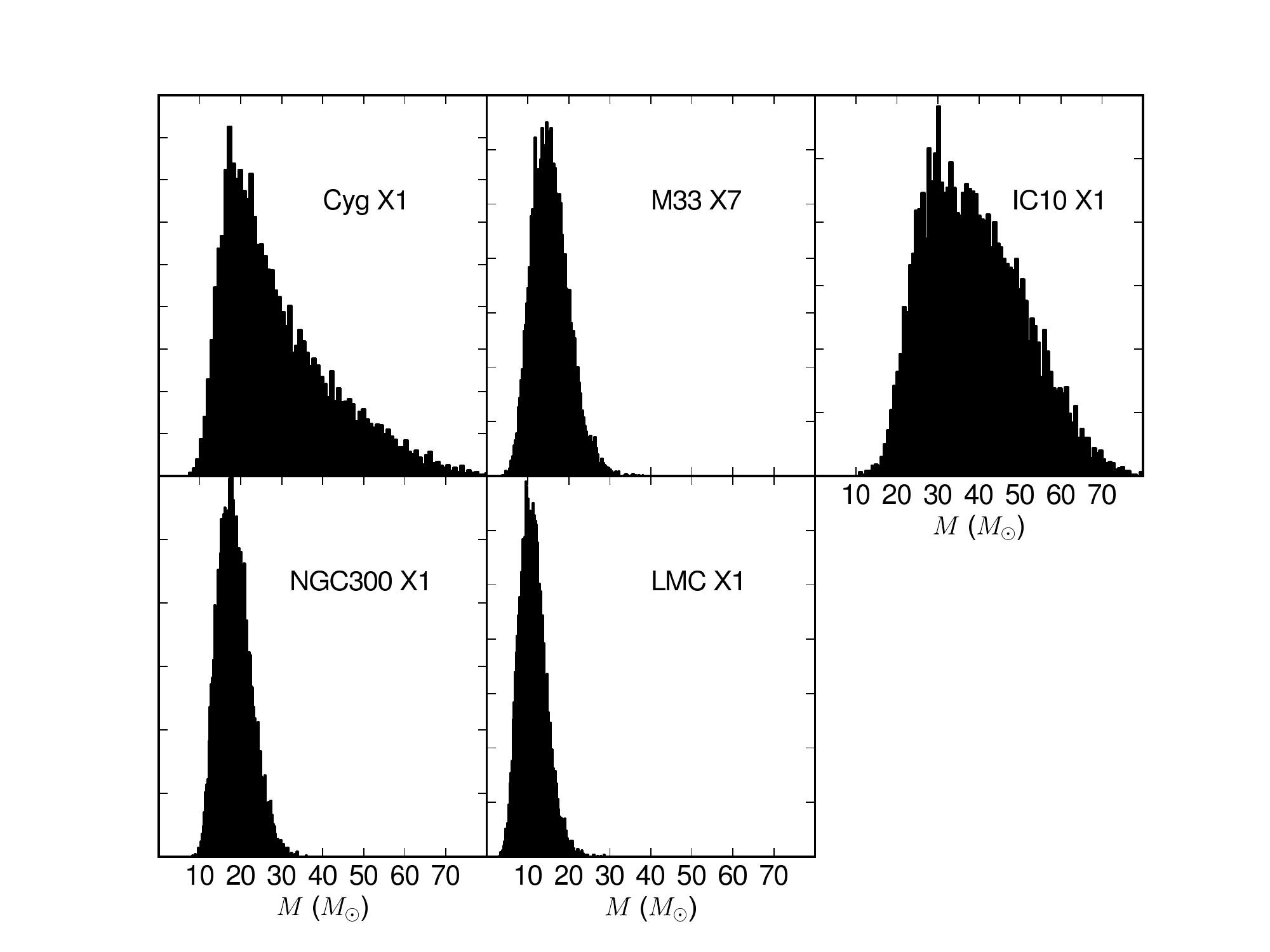}
  \end{center}
  \caption{\label{fig:high-masses} Mass distributions for the
    wind-fed, high-mass systems computed from the distributions on
    observed data in Table \ref{tab:sources} using Equation
    \eqref{eq:mass-function}.  (Similar to Figure
    \ref{fig:low-masses}.)  The asymmetry and long tails in these
    distributions are the result of the non-linear relationship
    between $M$, $f$, $q$, and $i$.}
\end{figure}

From these assumptions, we can generate probability distributions for
the true mass of the black hole given the observations and errors via
the Monte Carlo method: drawing samples of $f$, $q$, and $i$ from the
assumed distributions and computing the mass implied by Equation
\eqref{eq:mass-function} gives samples of $M$ from the distribution
induced by the relationship in Equation \eqref{eq:mass-function}.
Mass distributions generated in this way for the systems used in this
work are shown in Figure \ref{fig:low-masses} and Figure
\ref{fig:high-masses}.  Systems for which $i$ is poorly constrained
have broad ``tails'' on their mass distributions.  These mass
distributions constitute the ``observational data'' we will use in the
remainder of this paper.

\section{Statistical Analysis}
\label{sec:models}

In this section we describe the statistical analysis we will apply to
various models for the underlying mass distribution from which the
low-mass sample and the combined sample of X-ray binary systems in
Table \ref{tab:sources} were drawn.  The results of our analysis are
presented in Section \ref{sec:results}.

\subsection{Bayesian Inference}

The end result of our statistical analysis will be the probability
distribution for the parameters of each model implied by the data from
Section \ref{sec:systems} in combination with our prior assumptions
about the probability distribution for the parameters.  Bayes' rule
relates these quantities.  For a model with parameters $\vtheta$ in
the presence of data $d$, Bayes' rule states
\begin{equation}
  \label{eq:Bayes-rule}
  p(\vtheta | d) = \frac{p(d | \vtheta) p(\vtheta)}{p(d)}.
\end{equation}
Here, $p(\vtheta|d)$, called the posterior probability distribution
function, is the probability distribution for the parameters $\vtheta$
implied by the data $d$; $p(d|\vtheta)$, called the likelihood, is the
probability of observing data $d$ given that the model parameters are
$\vtheta$; $p(\vtheta)$, called the prior, reflects our estimate of the
probability of the various model parameters in the absence of any
data; and $p(d)$, called the evidence, is an overall normalizing
constant ensuring that
\begin{equation}
  \int d\theta\, p(\vtheta|d) = 1,
\end{equation}
whence
\begin{equation}
  \label{eq:evidence-def}
  p(d) = \int d\vtheta\, p(d|\vtheta) p(\vtheta).
\end{equation}

In our context, the data are the mass distributions given in Section
\ref{sec:systems}: $d = \{ p_i(M)| i = 1, 2, \ldots, 20 \}$.  We
assume that the measurements in Section \ref{sec:systems} are
independent, so the complete likelihood is given by a product of the
likelihoods for the individual measurements.  For a model with
parameters $\vtheta$ that predicts a mass distribution $p(M|\vtheta)$
for black holes, we have
\begin{equation}
  \label{eq:likelihood-def}
  p(d|\vtheta) = \prod_i \int dM\, p_i(M) p(M|\vtheta).
\end{equation}
That is, the likelihood of an observation is the average over the
individual mass distribution implied by the observation, $p_i(M)$, of
the probability for a black hole of that mass to exist according to
the model of the mass distribution, $p(M | \vtheta)$.  We approximate
the integrals as averages of $p(M|\vtheta)$ over the Monte Carlo mass
samples drawn from the distributions in Table \ref{tab:sources} (also
see Figures \ref{fig:low-masses} and \ref{fig:high-masses}):
\begin{equation}
  p(d|\vtheta) \approx \prod_i \frac{1}{N_i} \sum_{j = 1}^{N_i} p(M_{ij} | \vtheta),
\end{equation}
where $M_{ij}$ is the $j$th sample (out of a total $N_i$) from the
$i$th individual mass distribution.

Our calculation of the likelihood of each observation does not include
any attempt to account for selection effects in the observations.  We
simply assume (almost certainly incorrectly) that any black hole drawn
from the underlying mass distribution is equally likely to be
observed.  The results of \citet{Ozel2010} suggest that selection
effects are unlikely to significantly bias our analysis.

For a mass distribution with several parameters, $p(\vtheta | d)$
lives in a multi-dimensional space.  Previous works
\citep{Ozel2010,Bailyn1998} have considered models with only two
parameters; for such models evaluating $p(\vtheta|d)$ on a grid may be
a reliable method.  Many of our models for the underlying mass
distribution have three or more parameters.  Exploring the entirety of
these parameter spaces with a grid rapidly becomes prohibitive as the
number of parameters increases.  A more efficient way to explore the
distribution $p(\vtheta | d)$ is to use a Markov Chain Monte Carlo
(MCMC) method (see Appendix \ref{sec:mcmc}).  MCMC methods produce a
chain (sequence) of parameter samples, $\{ \vtheta_i \, | \, i = 1,
\ldots \}$, such that a particular parameter sample, $\vtheta$,
appears in the sequence with a frequency proportional to its posterior
probability, $p(\vtheta|d)$.  In this way, regions of parameter space
where $p(\vtheta|d)$ is large are sampled densely while regions where
$p(\vtheta|d)$ is small are effectively ignored.

Once we have a chain of samples from $p(\vtheta|d)$, the distribution
for any quantity of interest can be computed by evaluating it on each
sample in the chain and forming a histogram of these values.  For
example, to compute the one-dimensional distribution for a single
parameter obtained by integrating over all other dimensions in
parameter space, called the ``marginalized'' distribution, one plots
the histogram of the values of that parameter appearing in the chain.

\subsection{Priors}
\label{sec:priors}

An important part of any Bayesian analysis is the priors placed on the
parameters of the model.  The choice of priors can bias the results of
the analysis through both the shape and the range of prior support in
parameter space.  The prior should reflect the ``best guess'' for the
distribution of parameters before examining any of the data.  In the
absence of any information about the distribution of parameters, it is
best to choose a prior that is broad and uninformative to avoid biasing
the posterior as much as possible.

A prior that is independent of parameters, $\vtheta$, in some region,
called ``flat,'' results in a posterior that is proportional to the
likelihood (see Equation \eqref{eq:Bayes-rule}).  A flat prior does
not change the shape of the posterior.  However, the choice of a flat
prior is parameterization-dependent: a change of parameter from
$\vtheta$ to $\vtheta' = \vec{f}(\vtheta)$ can change a flat
distribution into one with non-trivial structure.  In this work, we
choose priors that are flat when the parameters are measured in
physical units.  In particular, for the log-normal model (Section
\ref{sec:log-normal}) the natural parameters for the distribution are
the mean, $\langle \log M \rangle$, and standard deviation,
$\sigma_{\log M}$, in $\log M$, but we choose priors that are flat in
$\langle M \rangle$ and $\sigma_M$.

The range of prior support can also affect the results of a Bayesian
analysis.  Because priors are normalized, prior support over a larger
region of parameter space results in a smaller prior probability at
each point.  Such ``wide'' priors are implicitly claiming that any
particular sub-region of parameter space is less likely than it would
be under a prior of the same shape but smaller support volume.  This
difference is important in model selection (Section
\ref{sec:model-selection}): when comparing two models with the same
likelihood, one with wide priors will seem less probable than one with
narrower priors.  Of course, priors should be wide enough to encompass
all regions of parameter space that have significant likelihood.  To
make the model comparison in Section \ref{sec:model-selection} fair,
we choose prior support in parameter space so that the allowed
parameter values for each model give distributions for which nearly
all the probability lies in the range $0 \leq M \leq 40 \Msun$.

\subsection{Parametric Models for the Black Hole Mass Distribution}
\label{sec:parametric-models}

Here we discuss the various parametric models of the underlying black
hole mass distribution considered in this paper.

\subsubsection{Power-Law Models}
\label{sec:power-law}

Many astrophysical distributions are power laws.  Let us assume that
the BH mass distribution is given by
\begin{equation}
  \label{eq:power-law-dist}
  p(M|\vtheta) = p(M|\{\Mmin, \Mmax, \alpha \}) =
  \begin{cases}
    A M^\alpha & \Mmin \leq m \leq \Mmax \\
    0 & \textnormal{otherwise}
  \end{cases}.
\end{equation}
The normalizing constant $A$ is 
\begin{equation}
  A = \frac{1+\alpha}{\Mmax^{1+\alpha} - \Mmin^{1+\alpha}}.
\end{equation}
We choose uniform priors on $\Mmin$ and $\Mmax \geq \Mmin$ between 0 and
$40 \Msun$, and uniform priors on the exponent $\alpha$ in a broad
range between $-15$ and $13$:
\begin{equation}
  p(\vtheta) = p(\{\Mmin, \Mmax, \alpha\}) = 
  \begin{cases}
    2 \frac{1}{40^2}\frac{1}{28} & 0 \leq \Mmin \leq \Mmax
    \leq 40, \quad -15 \leq \alpha \leq 13 \\
    0 & \textnormal{otherwise}
  \end{cases}.
\end{equation}

Our MCMC analysis output is a list of $\{\Mmin, \Mmax, \alpha\}$
values distributed according to the posterior 
\begin{equation}
  p(\vtheta|d) = p(\{\Mmin, \Mmax, \alpha\}|d) \propto p(d|\{\Mmin,
  \Mmax, \alpha\}) p(\{\Mmin, \Mmax, \alpha\}),
\end{equation}
with the likelihood $p(d|\{\Mmin, \Mmax, \alpha\})$ defined in
Equation \eqref{eq:likelihood-def}.  

\subsubsection{Decaying Exponential}
\label{sec:exponential}

\citet{Fryer2001} studied the relation between progenitor and remnant
mass in simulations of supernova explosions.  Combining this with the
mass function for supernova progenitors, they suggested that the
black-hole mass distribution may be well-represented by a decaying
exponential with a minimum mass: 
\begin{equation}
  \label{eq:exp-def}
  p(M|\vtheta) = p(M|\{\Mmin, M_0\}) = 
  \begin{cases}
    \frac{e^{\frac{\Mmin}{M_0}}}{M_0} \exp \left[ - \frac{M}{M_0}
    \right] & M \geq \Mmin \\
    0 & \textnormal{otherwise}
  \end{cases}.
\end{equation}
We choose a prior for this model where $\Mmin$ is uniform between 0
and $40 \Msun$.  For each $\Mmin$, we choose $M_0$ uniformly within a
range ensuring that $40\Msun$ is at least two scale masses above the
cutoff: $40\Msun \geq \Mmin + 2 M_0$.  This ensures that the majority
of the mass probability lies in the range $0 \leq M \leq 40\Msun$.
The resulting prior is
\begin{equation}
  p(\vtheta) = p(\{\Mmin, M_0\}) = 
  \begin{cases}
    \frac{4}{40^2} & 0 \leq \Mmin \leq 40, \quad 0 < M_0, \quad \Mmin
    + 2 M_0 \leq 40, \\
    0 & \textnormal{otherwise}
  \end{cases}
\end{equation}

\subsubsection{Gaussian and Two-Gaussian Models}
\label{sec:gaussian}

The mass distributions in Figure \ref{fig:low-masses} all peak in a
relatively narrow range near $\sim 10 \Msun$.  The prototypical
single-peaked probability distribution is a Gaussian:
\begin{equation}
  \label{eq:gaussian-def}
  p(M|\vtheta) = p(M|\{\mu, \sigma\}) = \frac{1}{\sigma \sqrt{2\pi}}
  \exp\left[ - \left(\frac{M - \mu}{\sqrt{2} \sigma} \right)^2 \right].
\end{equation}
We use a prior on the mean mass, $\mu$, and the standard deviation,
$\sigma$, that ensures that the majority of the mass distribution lies
below $40 \Msun$:
\begin{equation}
  \label{eq:gaussian-prior-def}
  p(\{\mu,\sigma\}) = 
  \begin{cases}
    \frac{8}{40^2} & 0 \leq \mu \leq 40, \quad \sigma \geq 0, \quad
    \mu + 2\sigma \leq 40 \\
    0 & \textnormal{otherwise}
  \end{cases},
\end{equation}
where both $\mu$ and $\sigma$ are measured in solar masses.  

Though we do not expect to find a second peak in the low-mass
distribution, we may find evidence of one when exploring the combined
low- and high-mass samples.  To look for a second peak in the
black-hole mass distribution, we use a two-Gaussian model:
\begin{multline}
  \label{eq:two-gaussian-def}
  p(M|\vtheta) = p(M|\{\mu_1, \mu_2, \sigma_1, \sigma_2, \alpha\}) = \\
  \frac{\alpha}{\sigma_1 \sqrt{2\pi}} \exp\left[ - \left( \frac{M -
        \mu_1}{\sqrt{2}\sigma_1} \right)^2 \right] + \frac{1-\alpha}{\sigma_2 \sqrt{2\pi}} \exp\left[ - \left( \frac{M -
        \mu_2}{\sqrt{2}\sigma_2} \right)^2 \right].
\end{multline}
The probability is a linear combination of two Gaussians with weights
$\alpha$ and $1-\alpha$.  We restrict $\mu_1 < \mu_2$ and also impose
combined conditions on $\mu_i$ and $\sigma_i$ that ensure that most of
the mass probability lies below $40 \Msun$ with the prior
\begin{equation}
  p(\{\mu_1, \mu_2, \sigma_1, \sigma_2, \alpha\}) = 
  \begin{cases}
    2 p(\{\mu_1, \sigma_1\}) p(\{\mu_2, \sigma_2\}) & \mu_1 \leq
    \mu_2, \quad 0 \leq \alpha \leq 1 \\
    0 & \textnormal{otherwise}
  \end{cases},
\end{equation}
where the single-Gaussian prior, $p(\{\mu_i, \sigma_i\})$, is defined
in Equation \eqref{eq:gaussian-prior-def}.

\subsubsection{Log Normal}
\label{sec:log-normal}

Many of the mass distributions for the systems in Figure
\ref{fig:low-masses} rise rapidly to a peak and then fall off more
slowly in a longer tail toward high masses.  So far, none of the
parameterized distributions we have discussed have this property.  In
this section, we consider a log-normal model for the underlying mass
distribution; the log-normal distribution has a rise to a peak with a
slower falloff in a long tail.  

The log-normal distribution gives $\log M$ a Gaussian distribution
with mean $\mu$ and standard deviation $\sigma$:
\begin{equation}
  \label{eq:log-normal-def}
  p(M|\vtheta) = p(M|\{\mu, \sigma \}) = \frac{1}{
    \sqrt{2\pi} M \sigma} \exp\left[ -\frac{\left(\log M - \mu\right)^2}{2 \sigma^2} \right].
\end{equation}
The parameters $\mu$ and $\sigma$ are dimensionless; the mean mass
$\langle M \rangle$ and mass standard deviation $\sigma_M$ are related
to $\mu$ and $\sigma$ by
\begin{eqnarray}
  \label{eq:avg-M}
  \langle M \rangle & = & \exp\left( \mu + \frac{1}{2} \sigma^2
  \right) \\
  \label{eq:sigma-M}
  \sigma_M & = & \langle M \rangle \sqrt{\exp\left( \sigma^2 \right) - 1}.
\end{eqnarray}
For a fair comparison with the other models, we impose a prior that is
flat in $\langle M \rangle$ and $\sigma_M$.  To ensure that most of
the probability in this model occurs for masses below $40 \Msun$, we
require $\langle M \rangle + 2 \sigma_M \leq 40$, resulting in a
prior
\begin{equation}
  p(\vtheta) = p(\{\mu,\sigma\}) = 
  \begin{cases}
    \frac{4}{40^2} \left| \frac{\partial \left(\langle M \rangle,
          \sigma_M \right)}{\partial \left( \mu, \sigma \right)}
    \right| & \sigma > 0, \quad \langle M \rangle + 2 \sigma_M \leq 40
    \\
    0 & \textnormal{otherwise}
  \end{cases},
\end{equation}
where 
\begin{equation}
  \left| \frac{\partial \left(\langle M \rangle,
          \sigma_M \right)}{\partial \left( \mu, \sigma \right)}
    \right| = \frac{\exp\left( 2 \left( \mu + \sigma^2 \right)\right)
      \sigma}{\sqrt{\exp\left( \sigma^2 \right) - 1}}
\end{equation}
is the determinant of the Jacobian of the map in Equations
\eqref{eq:avg-M} and \eqref{eq:sigma-M}.

\subsection{Non-Parametric Models for the Black Hole Mass Distribution}
\label{sec:non-parametric-models}

The previous subsection discussed models for the underlying black hole
mass distribution that assumed particular parameterized shapes for the
distribution.  In this subsection, we will discuss models that do not
assume a priori a shape for the black hole mass distribution.  The
fundamental non-parametric distribution in this section is a
histogram with some number of bins, $\Nbin$.  Such a distribution is
piecewise-constant in $M$.

One choice for representing such a histogram would be to fix the bin
locations, and allow the heights to vary.  With this approach, one
should be careful not to ``split'' features of the mass distribution
across more than one bin in order to avoid diluting the sensitivity to
such features; similarly, one should avoid including more than ``one''
feature in each bin.  The locations of the bins, then, are crucial.
An alternative representation of histogram mass distributions avoids
this difficulty.

We choose to represent a histogram mass distribution with $\Nbin$ bins
by allocating a fixed probability, $1/\Nbin$, to each bin.  The lower
and upper bounds for each bin are allowed to vary; when these are
close to each other (i.e.\ the bin is narrow), the distribution will
have a large value, and conversely when the bounds are far from each
other.  We assume that the non-zero region of the distribution is
contiguous, so we can represent the boundaries of the bins as a
non-decreasing array of masses, $w_0 \leq w_1 \leq \ldots \leq
w_{\Nbin}$, with $w_0$ the minimum and $w_{\Nbin}$ the maximum mass
for which the distribution has support.  This gives the distribution
\begin{equation}
  \label{eq:hist-def}
  p(M|\theta) = p(M|\{w_0, \ldots, w_{\Nbin}\}) = 
  \begin{cases}
    0 & M < w_0 \textnormal{ or } w_{\Nbin} \leq M \\
    \frac{1}{\Nbin} \frac{1}{w_{i+1} - w_i} & w_i \leq M < w_{i+1}
  \end{cases}.
\end{equation}

For priors on the histogram model with $\Nbin$ bins, we assume that
the bin boundaries are uniformly distributed between 0 and $40 \Msun$
subject only to the constraint that the boundaries are non-decreasing
from $w_0$ to $w_{\Nbin}$:
\begin{equation}
  p(\{w_0, \ldots, w_{\Nbin}\}) = 
  \begin{cases}
    \frac{\left(\Nbin+1\right)!}{40^{\Nbin+1}} & 0 \leq w_0 \leq w_1
    \leq \ldots \leq w_{\Nbin} \leq 40 \\
    0 & \textnormal{otherwise}
  \end{cases}.
\end{equation}

We consider histograms with up to five bins in this work.  We will see
that the evidence for the histogram models (see Sections
\ref{sec:model-selection}, \ref{sec:low-mass-model-selection}, and
\ref{sec:high-mass-model-selection}) from both the low-mass and
combined datasets is decreasing as the number of bins reaches five,
indicating that increasing the number of bins beyond five would not
sufficiently improve the fit to the mass distribution to compensate
for the extra parameter-space volume implied by the additional
parameters.

\subsection{Bayesian Model Selection}
\label{sec:model-selection}

In Sections \ref{sec:parametric-models} and
\ref{sec:non-parametric-models}, we discussed a series of models for
the underlying black hole mass distribution.  Our MCMC analysis will
provide the posterior distribution of the parameters within each
model, but does not tell us which models are more likely to correspond
to the actual distribution.  This model selection problem is the topic
of this section.

Consider a set of models, $\{M_i| i = 1, \ldots\}$, each with
corresponding parameters $\vtheta_i$.  Re-writing Equation
\eqref{eq:Bayes-rule} to be explicit about the assumption of a
particular model, we have
\begin{equation}
  p(\vtheta_i | d, M_i) = \frac{p(d|\vtheta_i, M_i) p(\vtheta_i | M_i)}{p(d|M_i)}.
\end{equation}
This gives the posterior probability of the parameters $\vtheta_i$ in
the context of model $M_i$.  But, the model itself can be regarded as
a discrete parameter in a larger ``super-model'' that encompasses all
the $M_i$.  The parameters for the super-model are $\{M_i,
\vtheta_i\}$: a choice of model and the corresponding parameter values
within that model.  Each point in the super-model parameter space is a
statement that, e.g., ``the underlying mass distribution is a
Gaussian, with parameters $\mu$ and $\sigma$,'' or ``the underlying
mass distribution is a triple-bin histogram with parameters $w_1$,
$w_2$, $w_3$, and $w_4$,'' or ....  The posterior probability of the
super-model parameters is given by Bayes' rule:
\begin{equation}
  \label{eq:bayes-explicit-model}
  p(\vtheta_i, M_i|d) = \frac{p(d|\vtheta_i, M_i) p(\vtheta_i |M_i) p(M_i)}{p(d)},
\end{equation}
where we have introduced the model prior $p(M_i)$, which represents
our estimate on the probability that model $M_i$ is correct in the
absence of the data $d$.  The normalizing evidence is now
\begin{equation}
  \label{eq:multi-model-evidence-def}
  p(d) = \sum_i \int d\vtheta_i\, p(d|\vtheta_i, M_i) p(\vtheta_i |M_i) p(M_i) = \sum_i
  p(d|M_i) p(M_i),
\end{equation}
writing the single-model evidence from Equation
\eqref{eq:evidence-def} as $p(d|M_i)$ to be explicit about the
dependence on the choice of model.

To compare the various models $M_i$, we are interested in the
marginalized posterior probability of $M_i$:
\begin{equation}
  \label{eq:model-posterior-def}
  p(M_i|d) \equiv \int d\vtheta_i\, p(\vtheta_i, M_i|d).
\end{equation}
This is the integral of the posterior over the entire parameter space
of model $M_i$.  The marginalized posterior probability of model $M_i$
can be re-written in terms of the single-model evidence, $p(d|M_i)$
(see Equations \eqref{eq:bayes-explicit-model} and
\eqref{eq:evidence-def}):
\begin{equation}
  \label{eq:model-evidence-def}
  p(M_i|d) = \int d\vtheta_i\, p(\vtheta_i, M_i|d) = \frac{p(M_i)}{p(d)} \int d\vtheta_i
  p(d|\vtheta_i,M_i) p(\vtheta_i|M_i) = \frac{p(d|M_i) p(M_i)}{p(d)}.
\end{equation}

Here and throughout, we assume that any of the models in Section
\ref{sec:models} are equally likely a priori, so the model priors are
equal:
\begin{equation}
  p(M_i) = \frac{1}{N_{\textnormal{model}}}.
\end{equation}

A powerful technique%
\footnote{We also attempted to compute $p(M_i|d)$ using two other
  methods: the well-known harmonic-mean estimator and the direct
  integration methods described in \citet{Weinberg2010}.  The harmonic
  mean is known to be very sensitive to outlying points in the MCMC in
  general, and we found this to be true in our specific application.
  The statistical properties of the direct integration algorithm from
  \citet{Weinberg2010} are less certain, but we found that it was
  quite noisy in our application compared to the reversible-jump MCMC.
  Due to the statistical noise in the other two methods, we use the
  results from our reversible jump MCMC analysis for model
  selection.} %
for computing $p(M_i|d)$ is the reversible-jump MCMC
\citep{Green1995}.  Reversible jump MCMC, discussed in more detail in
Appendix \ref{sec:reversible-jump-mcmc}, is a standard MCMC analysis
conducted in the super-model.  The result of a reversible jump MCMC is
a chain of samples, $\{ M_i, \vtheta_i\, | \, i = 1, \ldots \}$, from the
super-model parameter space.  The integral in Equation
\eqref{eq:model-evidence-def} can be estimated by counting the number
of times that a given model $M_i$ appears in the reversible jump MCMC
chain:
\begin{equation}
  p(M_i|d) = \int d\vtheta_i p(M_i, \vtheta_i|d) \approx \frac{N_i}{N},
\end{equation}
where $N_i$ is the number of MCMC samples that have discrete parameter
$M_i$, and $N$ is the total number of samples in the MCMC. 

Naively implemented reversible jump MCMCs can be very inefficient when
the posteriors for a model or models are strongly peaked.  In this
circumstance, a proposed MCMC jump into one of the peaked models is
unlikely to land on the peak by chance; since it is rare to propose a
jump into the important regions of parameter space of the peaked model
in a naive reversible jump MCMC, the output chain must be very long to
ensure that all models have been compared fairly.  We describe a new
algorithm in Appendix \ref{sec:reversible-jump-mcmc} that produces
very efficient jump proposals for a reversible jump MCMC by exploiting
the information about the model posteriors we have from the
single-model MCMC samples.  (See also \citet{Farr2010}.)  With this
algorithm, reasonable chain lengths can fairly compare all the models
under consideration.  We have used this algorithm to perform 10-way
reversible jump MCMCs to calculate the relative evidence for both the
parametric and non-parametric models in this study.  These results
appear in Section \ref{sec:results}.

\section{Results}
\label{sec:results}

In this section we discuss the results of our MCMC analysis of the
posterior distributions of parameters for the models in Sections
\ref{sec:parametric-models} and \ref{sec:non-parametric-models}.  We
also discuss model selection results.  The results in Section
\ref{sec:results-low-mass} apply to the low-mass sample of
systems, while those of Section \ref{sec:high-mass} apply to the
combined sample of systems.

\subsection{Low-Mass Systems}
\label{sec:results-low-mass}

Table \ref{tab:low-mass-parametric} gives quantiles of the
marginalized parameter distributions of the parametric models implied
by the low-mass data.  Table \ref{tab:low-mass-non-parametric} gives
the quantiles of the histogram bin boundaries in the non-parametric
analysis implied by the low-mass data.

\begin{table}
  \begin{center}
    \begin{tabular}{|l|c|c|c|c|c|c|}
      \hline
      Model & Parameter & 5\% & 15\% & 50\% & 85\% & 95\% \\
      \hline \hline
      Power Law (Equation \eqref{eq:power-law-dist}) & $\Mmin$ & 
      1.2786 &  4.1831 &  6.1001 &  6.5011 &  6.6250 \\
      \hline
       & $\Mmax$ & 8.5578 &  8.9214 & 23.3274 & 36.0002 & 38.8113 \\
       \hline
       & $\alpha$ & -12.4191 & -10.1894 & -6.3861 &  2.8476 &  5.6954 \\
       \hline \hline
       Exponential (Equation \eqref{eq:exp-def}) & $\Mmin$ & 
       5.0185 &  5.4439 &  6.0313 &  6.3785 &  6.5316 \\
       \hline
       & $M_0$ & 0.7796 &  0.9971 & 1.5516 &  2.4635 &  3.2518 \\
       \hline \hline
       Gaussian (Equation \eqref{eq:gaussian-def}) & $\mu$ & 
       6.6349 &  6.9130 &  7.3475 & 7.7845 & 8.0798 \\
       \hline
       & $\sigma$ & 0.7478 &  0.9050  & 1.2500 &  1.7335 & 2.1134 \\
       \hline \hline
       Two Gaussian (Equation \eqref{eq:two-gaussian-def}) & $\mu_1$ & 
       5.4506 &  6.3877 &  7.1514 &  7.6728  & 7.9803 \\
       \hline
       & $\mu_2$ & 7.2355 &  7.7387 & 12.3986 & 25.2456 & 31.4216 \\
       \hline
       & $\sigma_1$ & 0.3758 &  0.7626 &  1.2104 &  1.7981 &  2.3065 \\
       \hline
       & $\sigma_2$ & 0.2048 & 0.6421 & 1.9182 &  5.2757  & 7.2625 \\
       \hline
       & $\alpha$ & 0.0983 &  0.3526 & 0.8871 &  0.9792 &  0.9936 \\
       \hline \hline
       Log Normal (Equation \eqref{eq:log-normal-def}) & $\langle M \rangle$ & 
       6.7619 &  7.0122 &  7.4336  &  7.9159  &  8.2942 \\
       \hline 
       & $\sigma_M$ & 0.7292  &  0.8920  & 1.2704  &  1.8695  &  2.4069 \\
       \hline
    \end{tabular}
  \end{center}
  \caption{\label{tab:low-mass-parametric} Quantiles of the
    marginalized distribution for each of the parameters in the models
    discussed in Section \ref{sec:parametric-models} implied by the low-mass data.  We indicate
    the 5\%, 15\%, 50\% (median), 85\%, and 95\% quantiles.  The
    marginalized distribution can be misleading when there are strong
    correlations between variables.  For example, while the
    marginalized distributions for the power law parameters are quite
    broad, the distribution of mass distributions implied by the power
    law MCMC samples is similar to the other models.  This occurs in
    spite of the broad marginalized distributions because of the
    correlations between the slope and limits of the power law
    discussed in Section \ref{sec:power-law}.}
\end{table}

\begin{figure}
  \begin{center}
    \plotone{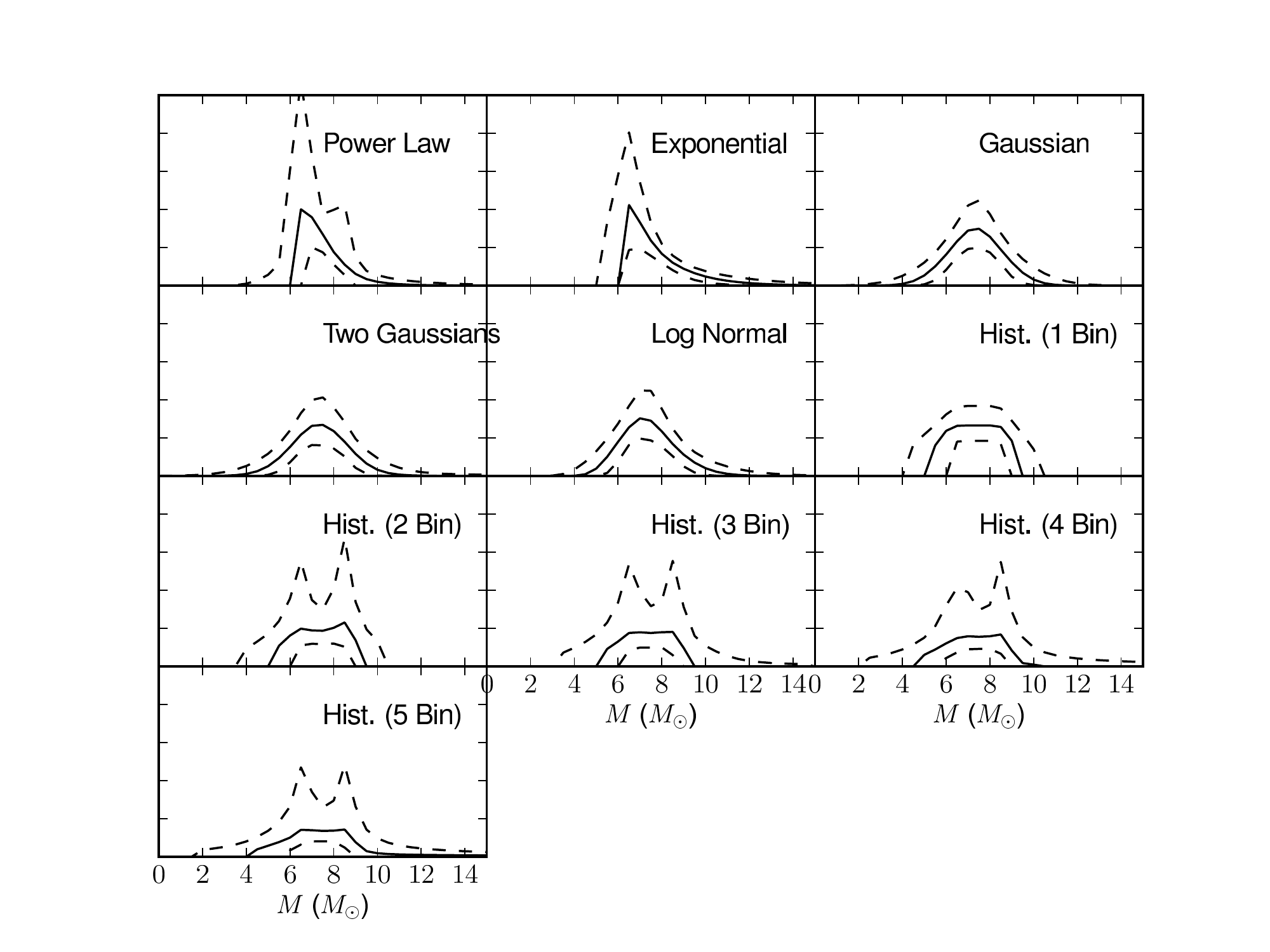}
  \end{center}
  \caption{\label{fig:dists} The median (solid line), 10\% (lower
    dashed line) and 90\% (upper dashed line) values of the black hole
    mass distribution, $p(M|\theta)$, at various masses implied by the
    posterior $p(\theta|d)$ for the models discussed in Sections
    \ref{sec:parametric-models} and \ref{sec:non-parametric-models}.
    These distributions use only the 15 low-mass observations in Table
    \ref{tab:sources} (the combined sample is analyzed in Section
    \ref{sec:high-mass}).  Note that these ``distributions of
    distributions'' are not necessarily normalized, and need not be
    shaped like the underlying model distributions.}
\end{figure}

Recall that each MCMC sample in our analysis gives the parameters for
a model of the black hole mass distribution.  The chain of samples of
parameters for a particular model gives us a distribution of black
hole mass distributions.  Figure \ref{fig:dists} gives a sense of the
shape and range of the distributions of black hole mass distributions
that result from our MCMC analysis.  In Figure \ref{fig:dists} we plot
the median, 10\% and 90\% values of the black hole mass distributions
that result from the MCMC chains.  Because the choice of parameters
that gives, for example, the median distribution value at one mass
need not give the median distribution at another mass, these curves do
not necessarily look like the underlying model for the mass
distribution.  For the same reason, they are not necessarily
normalized.

\subsubsection{Power Law}

\begin{figure}
  \begin{center}
    \plotone{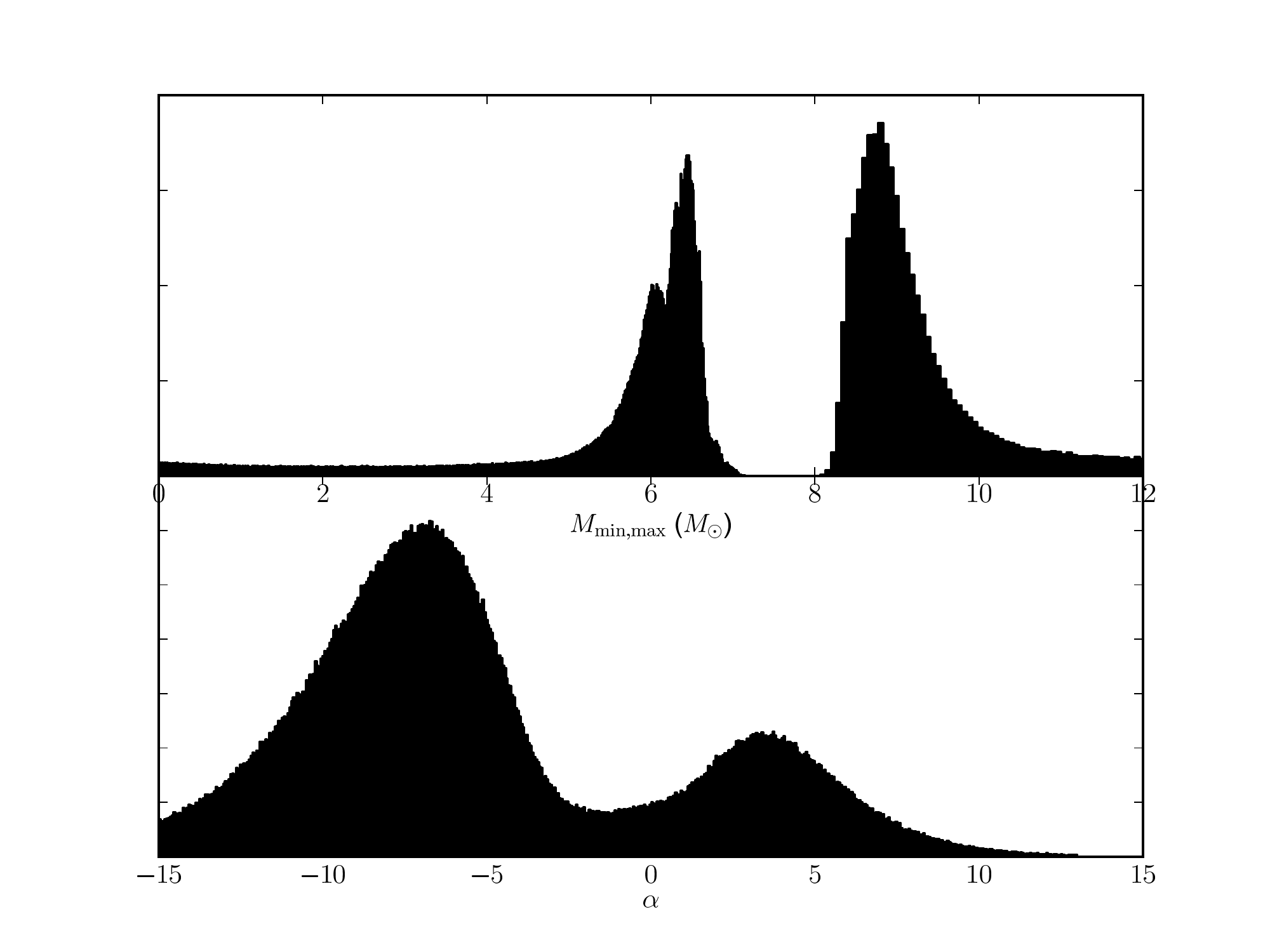}
  \end{center}
  \caption{\label{fig:power-law} Histograms of the marginalized
    distribution for the three parameters $\Mmin$ (top, left), $\Mmax$
    (top, right), and $\alpha$ (bottom) from the power-law model.  The
    marginalized distribution for $\alpha$ is broad, with $-11.8 <
    \alpha < 6.8$ enclosing 90\% of the probability.  We have
    $p(\alpha < 0) = 0.6$; the median value is $\alpha = -3.35$.  The
    broad distribution for $\alpha$ (and the other parameters) is due
    to correlations between the parameters discussed in the main text;
    see Figure \ref{fig:power-law-2D}.}
\end{figure}

In Figure \ref{fig:power-law} , we display a histogram of the
resulting samples in each of the parameters $\Mmin$, $\Mmax$, and
$\alpha$ for the power law model (see Equation
\eqref{eq:power-law-dist}); this represents the one-dimensional
``marginalized'' distribution
\begin{equation}
  \label{eq:alpha-pdf}
  p(\alpha|d) = \int d\Mmin\, d\Mmax\, p(\{\Mmin, \Mmax, \alpha\}|d),
\end{equation}
and similarly for $\Mmin$ and $\Mmax$.

\begin{figure}
  \begin{center}
    \plotone{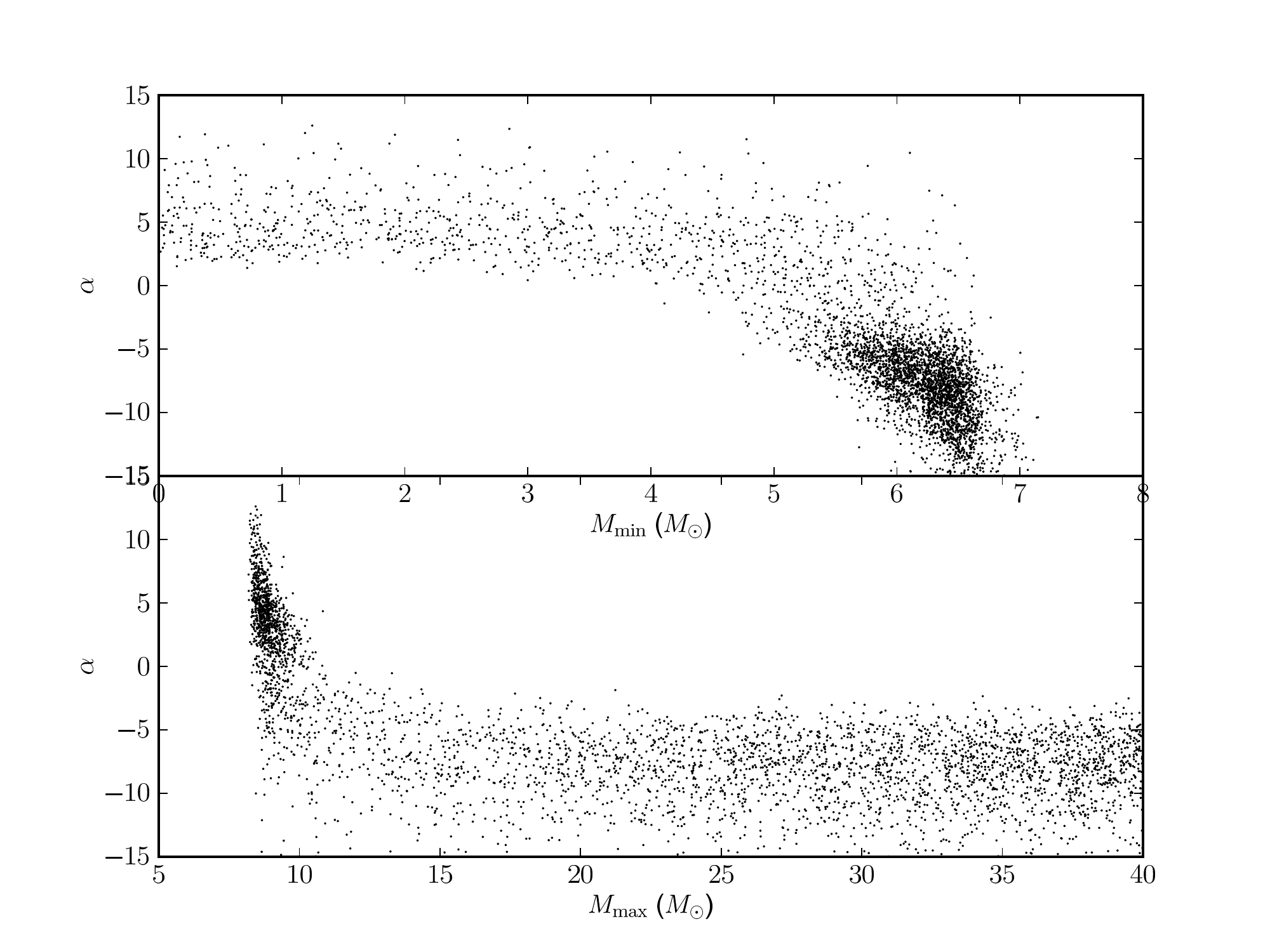}
  \end{center}
  \caption{\label{fig:power-law-2D} MCMC samples in the $\Mmin,
    \alpha$ (top) and $\Mmax, \alpha$ (bottom) planes for the
    power-law model discussed in Section \ref{sec:power-law}.  The
    correlations between $\alpha$ and the power law bounds discussed
    in the text are apparent: when $\alpha$ is positive, the mass
    distribution slopes upward and $\Mmax$ is constrained to be near
    the maximum observed mass while $\Mmin$ is unconstrained.  When
    $\alpha$ is negative, the mass distribution slopes down and
    $\Mmin$ is constrained to be near the lowest mass observed, while
    $\Mmax$ is unconstrained. }
\end{figure}

The marginalized distribution for $\alpha$ is broad, with
\begin{equation}
  -11.8 < \alpha < 6.8
\end{equation}
enclosing 90\% of the probability (excluding 5\% on each side).  We
have $p(\alpha < 0) = 0.6$.  The median value is $\alpha = -3.35$.
The broadness of the marginalized distribution for $\alpha$ comes from
the need to match the relatively narrow range in mass of the
low-mass systems.  When $\alpha$ is negative, the resulting mass
distribution slopes down; $\Mmin$ is constrained to be near the lowest
mass of the observed black holes, while $\Mmax$ is essentially
irrelevant.  Conversely, when $\alpha$ is positive and the mass
distribution slopes up, $\Mmax$ must be close to the largest mass
observed, while $\Mmin$ is essentially irrelevant.  Figure
\ref{fig:power-law-2D} illustrates this effect, showing the
correlations between $\alpha$ and $\Mmin$ and $\alpha$ and $\Mmax$.
When we include the high-mass systems in the analysis, the long tail
will eliminate this effect by bringing both $\Mmin$ and $\Mmax$ into
play for all values of $\alpha$.

\subsubsection{Decaying Exponential}

\begin{figure}
  \begin{center}
    \plotone{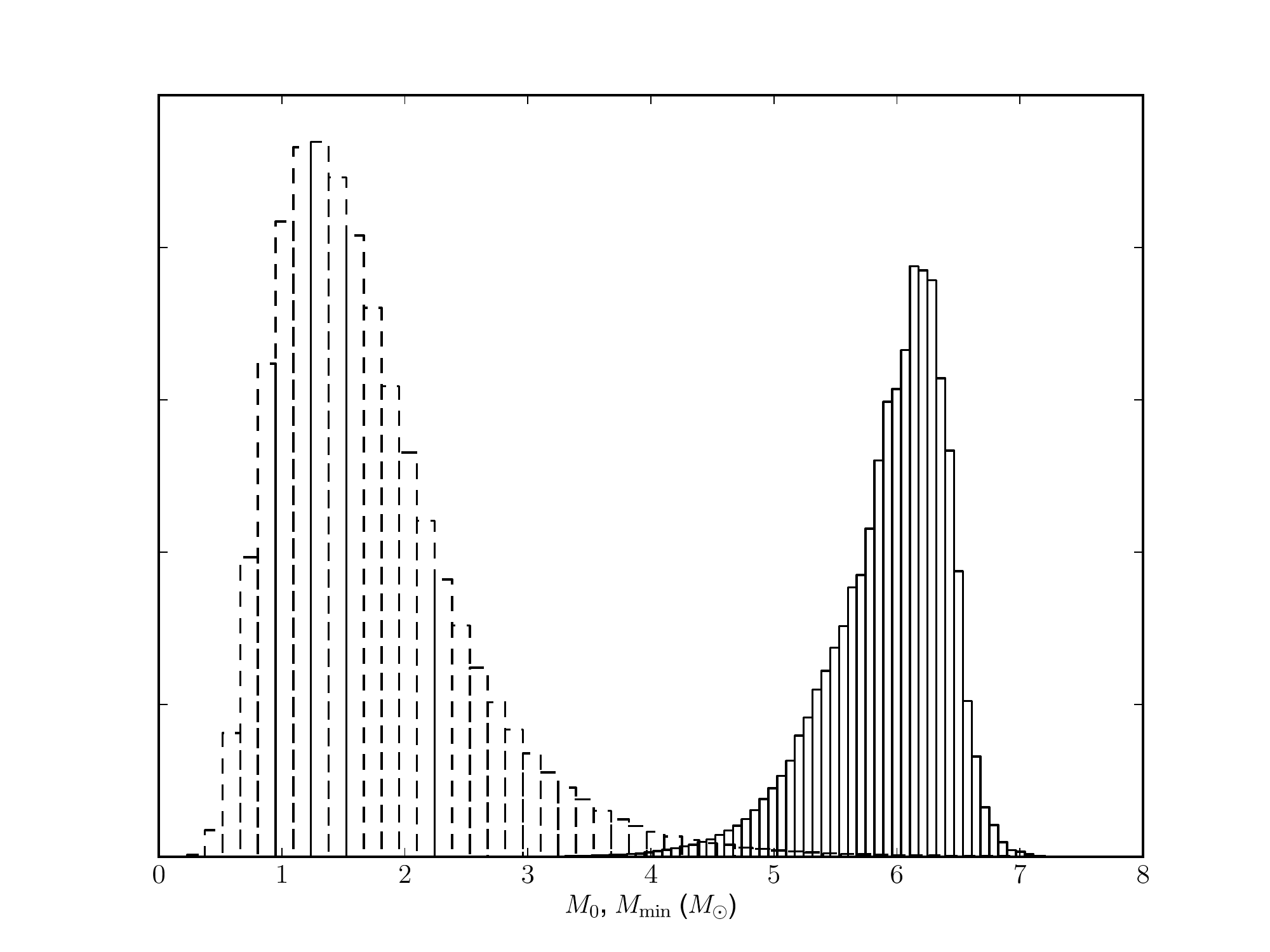}
  \end{center}
  \caption{\label{fig:exp-marginal} The distribution of scale masses,
    $M_0$ (dashed histogram), and minimum masses, $\Mmin$ (solid
    histogram), both measured in units of a solar mass for the
    exponential underlying mass distribution defined in Equation
    \eqref{eq:exp-def}.  The median scale mass is $M_0 = 1.55$, and
    $0.78 \leq M_0 \leq 3.25$ with 90\% confidence.}
\end{figure}

\begin{figure}
  \begin{center}
    \plotone{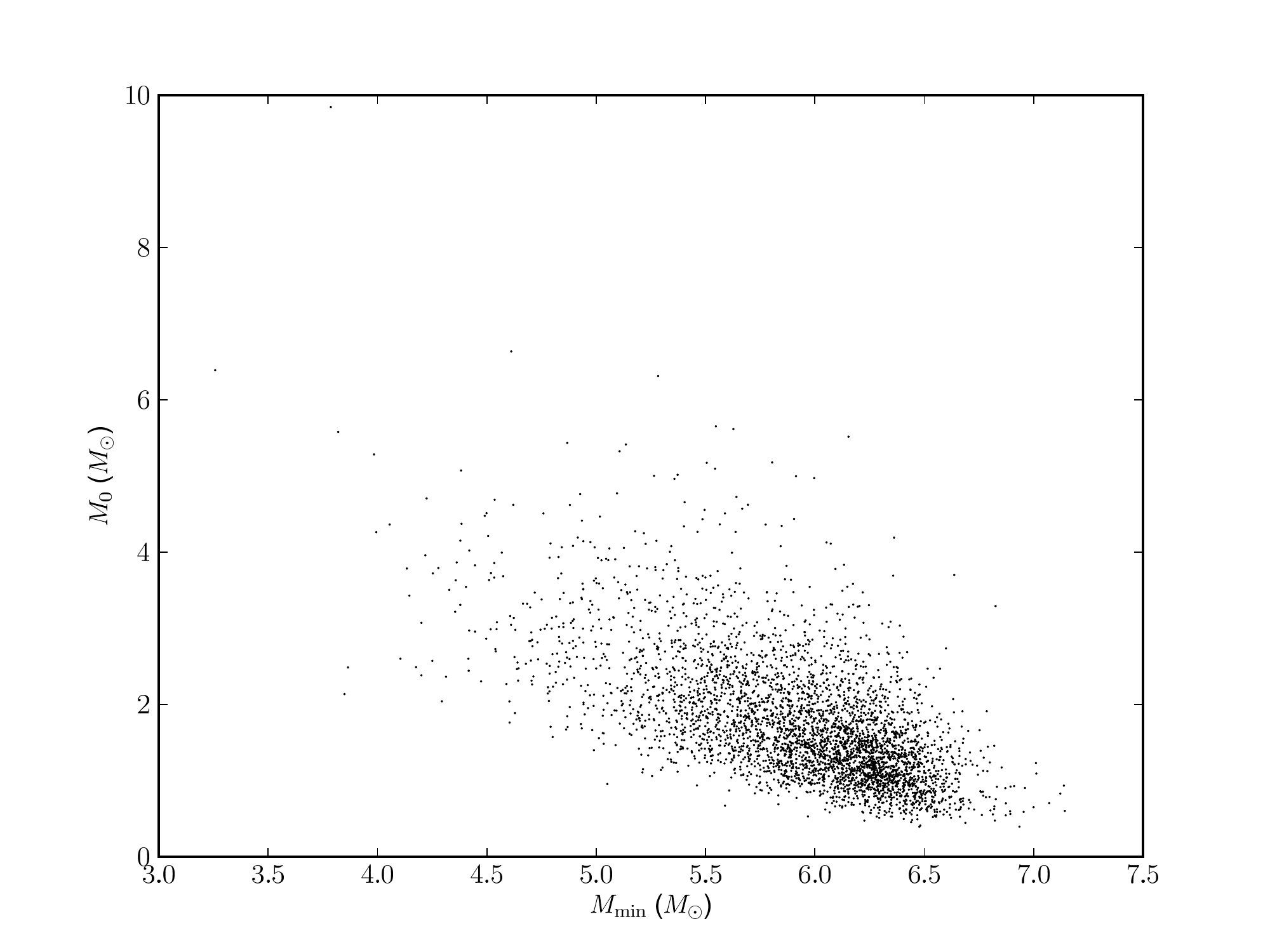}
  \end{center}
  \caption{\label{fig:exp-2D} The MCMC samples in the $\Mmin$, $M_0$
    plane for the decaying exponential underlying mass distribution
    model.  The slight correlation between smaller $\Mmin$ and larger
    $M_0$ is driven by the need to widen the mass distribution to
    encompass the peak of the measurements in Figure
    \ref{fig:low-masses} when the minimum mass decreases.}
\end{figure}

Figure \ref{fig:exp-marginal} displays the marginalized posterior
distribution for the scale mass of the exponential, $M_0$, and the
cutoff mass, $\Mmin$ (see Equation \ref{eq:exp-def}).  The median
scale mass is $M_0 = 1.55$, and $0.78 \leq M_0 \leq 3.25$ with 90\%
confidence.  This model was one of those considered by
\citet{Ozel2010}, whose results ($M_0 \sim 1.5$ and $\Mmin \sim 6.5$)
are broadly consistent with ours.  Figure \ref{fig:exp-2D} displays
the MCMC samples in the $\Mmin$, $M_0$ plane for this model.  There is
a small correlation between smaller $\Mmin$ and larger $M_0$, which is
driven by the need to widen the distribution to encompass the peak of
the mass measurements in Figure \ref{fig:low-masses} when the minimum
mass is smaller.

\subsubsection{Gaussian}

\begin{figure}
  \begin{center}
    \plotone{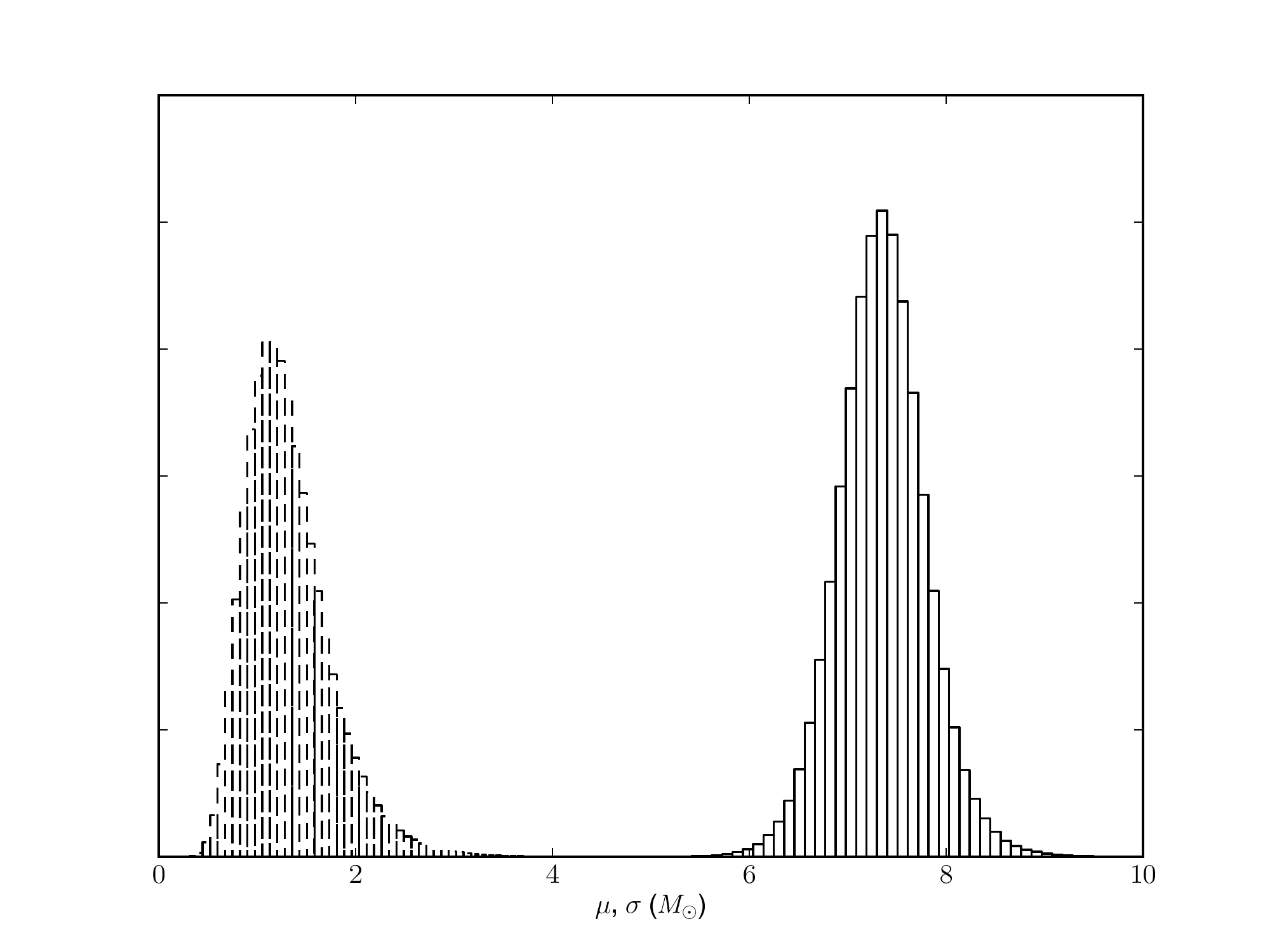}
  \end{center}
  \caption{\label{fig:gaussian} Marginalized posterior distributions
    for the mean, $\mu$ (solid histogram), and standard deviation,
    $\sigma$ (dashed histogram), both in solar masses for the Gaussian
    underlying mass distribution defined in Equation
    \eqref{eq:gaussian-def}.  The peak of the Gaussian, $\mu$, is
    constrained in $6.63 \leq \mu \leq 8.08$ with 90\% confidence.}
\end{figure}

Figure \ref{fig:gaussian} shows the resulting marginalized
distributions for the parameters $\mu$ and $\sigma$.  We constrain the
peak of the Gaussian between $6.63 \leq \mu \leq 8.08$ with 90\%
confidence.  This model also appeared in \citet{Ozel2010}; they found
$\mu \sim 7.8$ and $\sigma \sim 1.2$, consistent with our results
here.

\subsubsection{Two Gaussian}

\begin{figure}
  \begin{center}
    \plotone{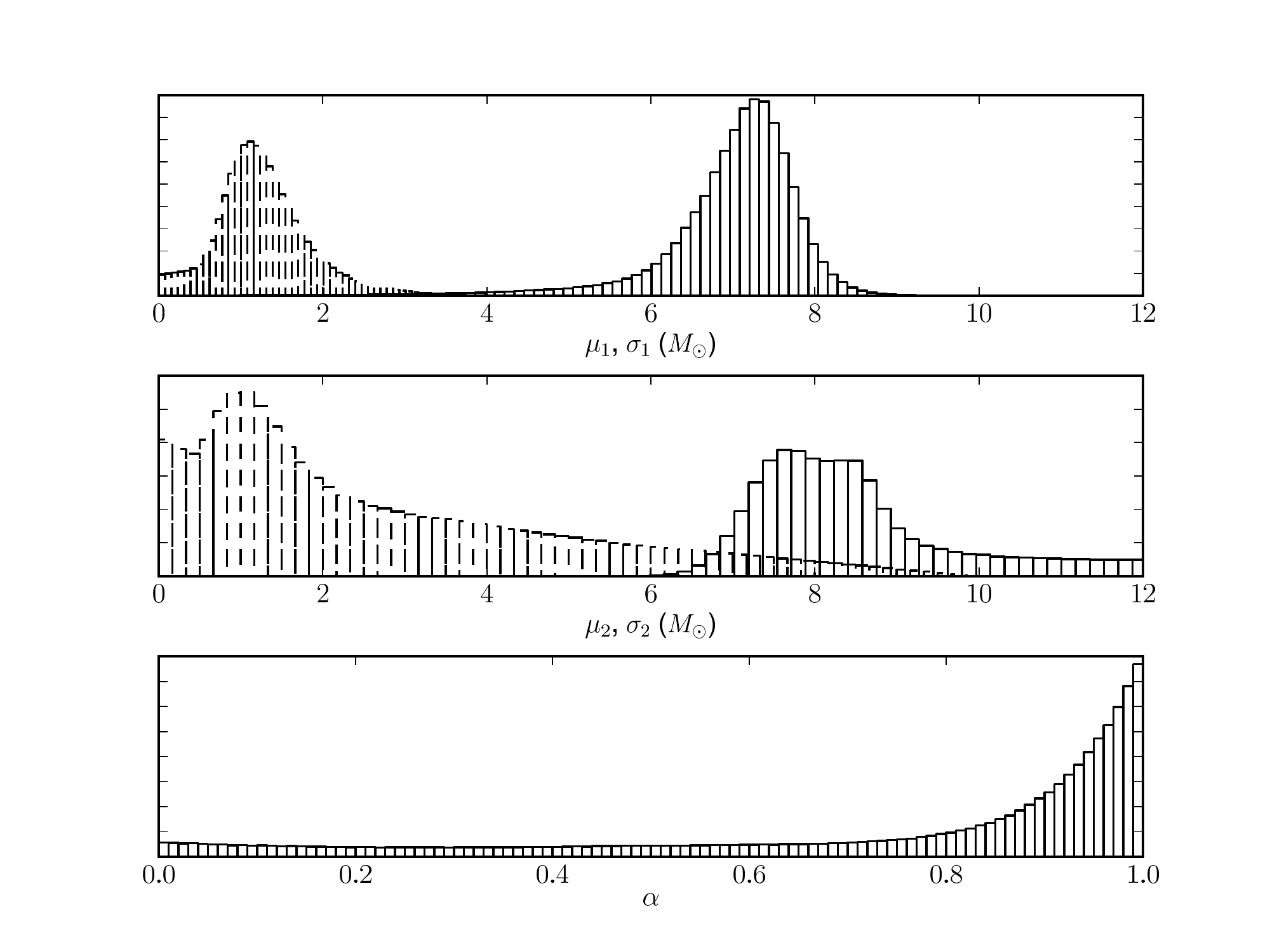}
  \end{center}
  \caption{\label{fig:two-gaussian} The marginal distributions for the
    five parameters of the two-Gaussian model.  The top panel is
    $\mu_1$ (solid histogram) and $\sigma_1$ (dashed histogram), the
    middle panel is $\mu_2$ (solid histogram) and $\sigma_2$ (dashed
    histogram), and the bottom panel is $\alpha$. We have $\alpha >
    0.8$ with 62\% probability, favoring the first of the two
    Gaussians.  The distributions for $\mu_1$ and $\sigma_1$ are
    similar to those of the single Gaussian model displayed in Figure
    \ref{fig:gaussian}; the second Gaussian's parameter distributions
    are much broader (recall that we constrain $\mu_2 > \mu_1$).  The
    second Gaussian is attempting to fit the tail of the mass samples.
    The extra degrees of freedom in the distribution from the second
    Gaussian do not provide enough extra fitting power to compensate
    for the increase in parmeter space, however: the two-Gaussian
    model is disfavored relative to the single Gaussian by a factor of
    $4.7$ on this dataset (see Sections \ref{sec:model-selection} and
    \ref{sec:low-mass-model-selection} for discussion).}
\end{figure}

Figure \ref{fig:two-gaussian} shows the marginalized distributions for
the two-Gaussian model parameters from our MCMC runs.  We find $\alpha
> 0.8$ with 62\% probability, clearly favoring the Gaussian with
smaller mean.  The distributions for $\mu_1$ and $\sigma_1$ are
similar to those of the single Gaussian displayed in Figure
\ref{fig:gaussian}, indicating that this Gaussian is centered around
the peaks of the low-mass distributions.  The second Gaussian's
parameter distributions are much broader.  The second Gaussian appears
to be sampling the tail of the mass samples.  In spite of the extra
degrees of freedom in this model, we find that this model is strongly
disfavored relative to the single-Gaussian model for this dataset:
$p(\textnormal{Gaussian}|d) / p(\textnormal{Two Gaussian}|d) \simeq
4.7$ (see Sections \ref{sec:model-selection} and
\ref{sec:low-mass-model-selection} for discussion).

\subsubsection{Log Normal}

\begin{figure}
  \begin{center}
    \plotone{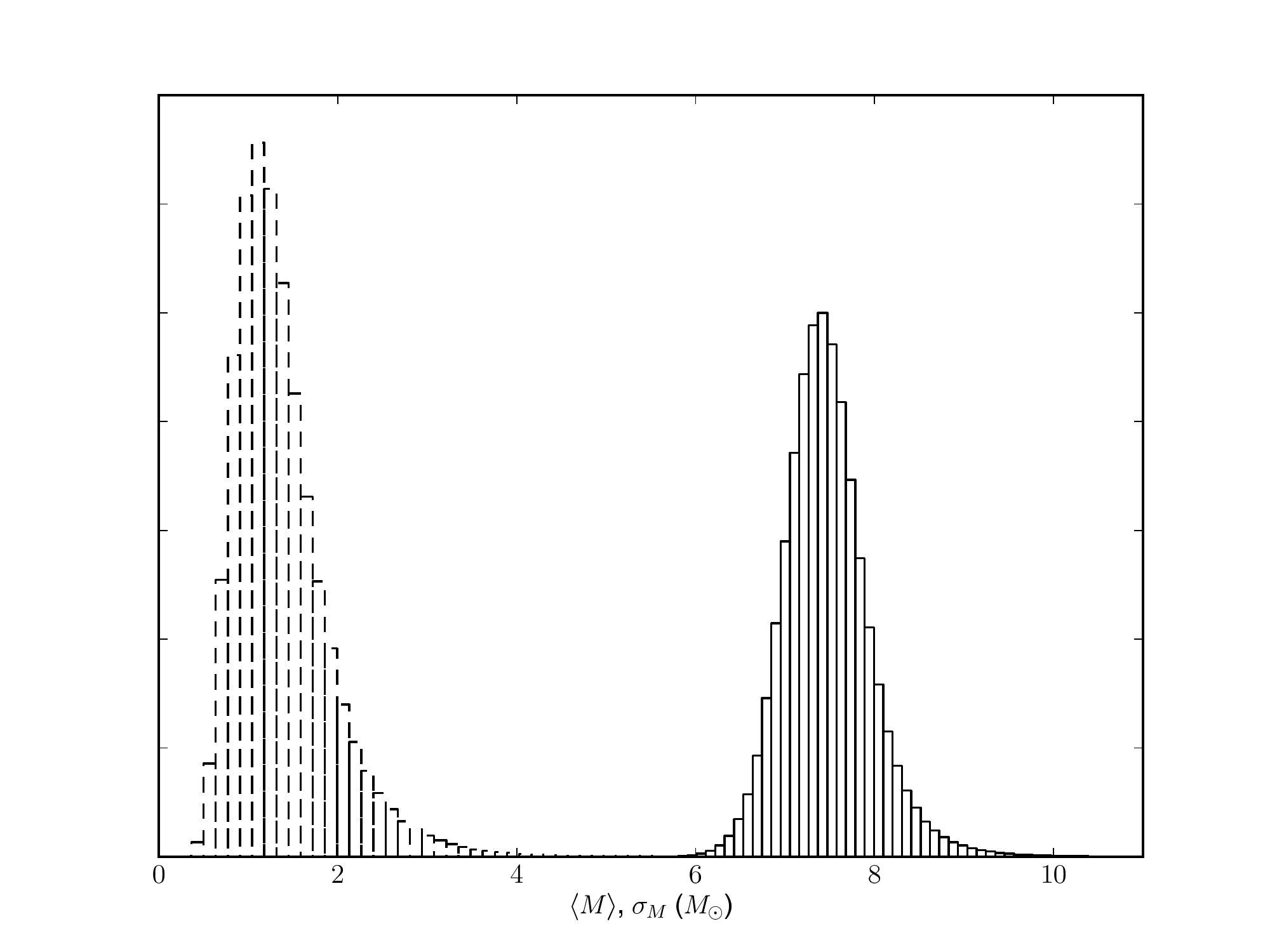}
  \end{center}
  \caption{\label{fig:log-normal} Marginalized distributions of the
    mean mass, $\langle M \rangle$ (solid histogram), and standard
    deviation of the mass, $\sigma_M$ (dashed histogram), for the
    log-normal model in Section \ref{sec:log-normal}.  The
    distributions are similar to the distributions of $\mu$ and
    $\sigma$ in the Gaussian model of Section \ref{sec:gaussian}.}
\end{figure}

The marginal distributions for $\langle M \rangle$ and $\sigma_M$
appear in Figure \ref{fig:log-normal}.  The distributions are similar
to those for $\mu$ and $\sigma$ from the Gaussian model in Section
\ref{sec:gaussian}.

\subsubsection{Histogram Models}

The median values of the histogram mass distributions that result from
the MCMC samples of the posterior distribution for the $w_i$
parameters for one-, two-, three-, four-, and five-bin histogram
models are shown in Figure \ref{fig:dists}.  Table
\ref{tab:low-mass-non-parametric} gives quantiles of the marginalized
bin boundary distributions for the histogram models.

\begin{table}
  \begin{center}
    \begin{tabular}{|c|c|c|c|c|c|c|}
      \hline
      Bins & Boundary & 5\% & 15\% & 50\% & 85\% & 95\% \\
      \hline \hline
      1 & $w_0$ & 3.94488 & 4.55603 & 5.43333 & 6.02557 & 6.29749 \\
      \hline
        & $w_1$ & 8.50844 & 8.69262 & 9.11784 & 9.83477 & 10.5128 \\
      \hline \hline
      2 & $w_0$ & 3.3426 & 4.2047 & 5.39132 & 6.18413 & 6.47553 \\
      \hline
        & $w_1$ & 6.41972 & 6.72605 & 7.43421 & 8.2489 & 8.52885 \\
      \hline
        & $w_2$ & 8.46161 & 8.65077 & 9.12694 & 10.1113 & 11.2595 \\
      \hline \hline
      3 & $w_0$ & 2.18176 & 3.54345 & 5.16094 & 6.16473 & 6.44697 \\
      \hline
        & $w_1$ & 5.68876 & 6.14223 & 6.68829 & 7.38725 & 8.04235 \\
      \hline
        & $w_2$ & 6.8297 & 7.22718 & 8.1451 & 8.7512 & 9.27296 \\
      \hline
        & $w_3$ & 8.44307 & 8.67362 & 9.25718 & 12.1688 & 21.92 \\
      \hline \hline
      4 & $w_0$ & 1.32131 & 2.7934 & 4.66156 & 5.78459 & 6.17946 \\
      \hline
        & $w_1$ & 5.20112 & 5.77331 & 6.42501 & 6.98427 & 7.44584 \\
      \hline
        & $w_2$ & 6.41805 & 6.73535 & 7.43826 & 8.32958 & 8.64212 \\
      \hline
        & $w_3$ & 7.40302 & 7.95608 & 8.58976 & 9.33897 & 10.3992 \\
      \hline
        & $w_4$ & 8.56724 & 8.8059 & 10.2451 & 24.3573 & 34.2423 \\
      \hline \hline
      5 & $w_0$ & 0.9392 & 2.28789 & 4.33389 & 5.7012 & 6.21166 \\
      \hline
        & $w_1$ & 4.69778 & 5.44302 & 6.26575 & 6.76407 & 7.14427 \\
      \hline
        & $w_2$ & 6.1388 & 6.47155 & 7.00606 & 7.97325 & 8.38259 \\
      \hline
        & $w_3$ & 6.82058 & 7.28677 & 8.22514 & 8.81555 & 9.41012 \\
      \hline
        & $w_4$ & 8.02335 & 8.36993 & 8.94879 & 11.3206 & 17.3349 \\
      \hline
        & $w_5$ & 8.7112 & 9.25208 & 16.2059 & 31.897 & 37.2738 \\
      \hline
    \end{tabular}
  \end{center}
  \caption{\label{tab:low-mass-non-parametric} The 5\%, 15\%, 50\%
    (median), 85\%, and 95\% quantiles for the bin boundaries in the
    one- through five-bin histogram models discussed in Section
    \ref{sec:non-parametric-models}.}
\end{table}

As the number of bins increases, the models are better able to capture
features of the mass distribution, but we find that the one-bin
histogram is the most probable of the histogram models for the
low-mass data (see Section \ref{sec:low-mass-model-selection} for
discussion).  This occurs because the extra fitting power does not
sufficiently improve the fit to compensate for the vastly larger
parameter space of the models with more bins.

\subsubsection{Model Selection for the Low-Mass Sample}
\label{sec:low-mass-model-selection}
\begin{figure}
  \begin{center}
    \plotone{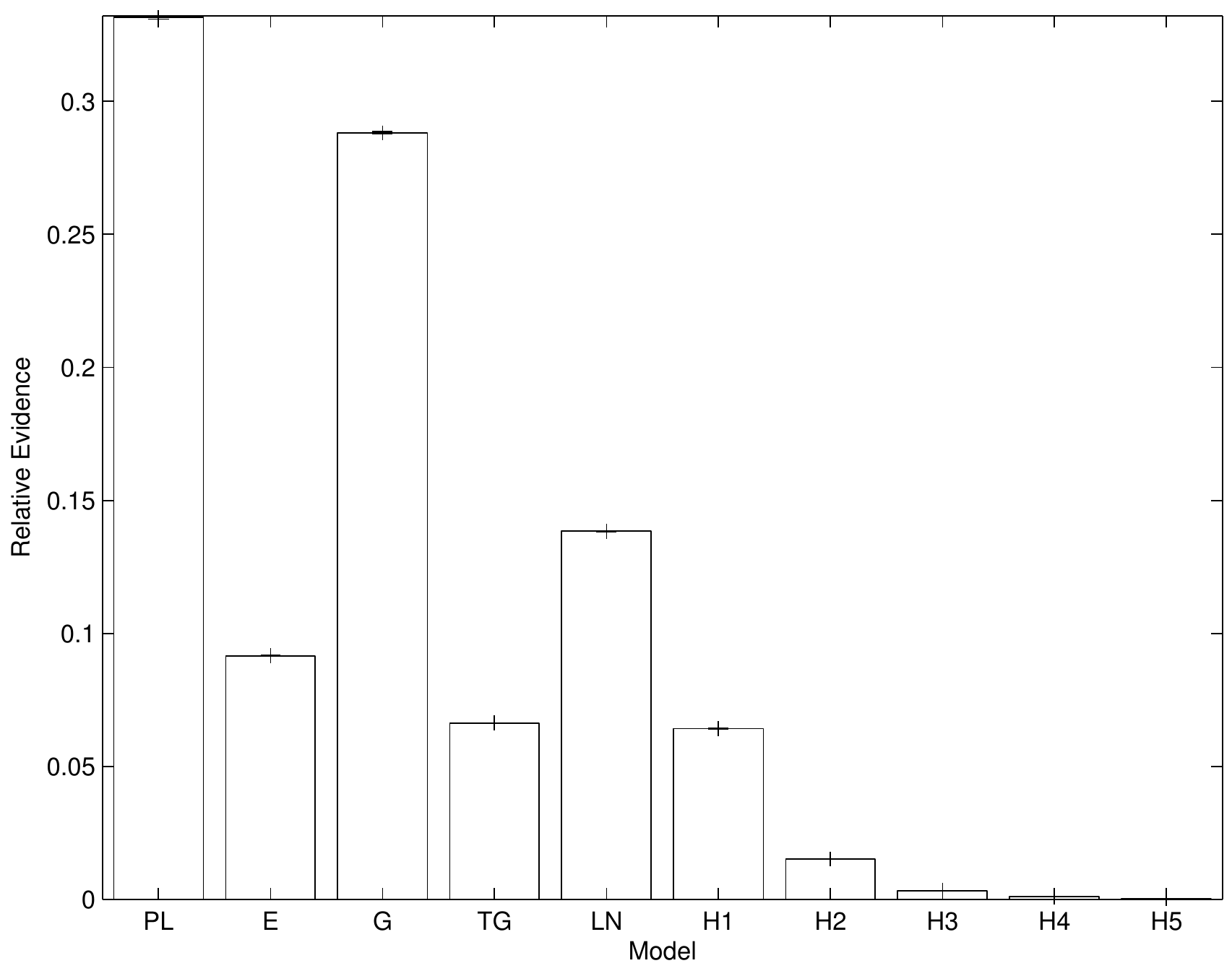}
  \end{center}
  \caption{\label{fig:rj} The relative probability of the models
    discussed in Section \ref{sec:models} as computed using the
    reversible-jump MCMC with the efficient jump proposal algorithm
    described in Section \ref{sec:reversible-jump-mcmc}.  (See also
    Table \ref{tab:rj}.)  In increasing order along the $x$-axis, the
    models are the power-law of Section \ref{sec:power-law} (PL), the
    decaying exponential of Section \ref{sec:exponential} (E), the
    single Gaussian of Section \ref{sec:gaussian} (G), the double
    Gaussian of Section \ref{sec:gaussian} (TG), and the one-, two-,
    three-, four-, and five-bin histogram models of Section
    \ref{sec:non-parametric-models} (H1, H2, H3, H4, H5,
    respectively).  The average of 500 independent reversible-jump
    MCMCs is plotted, along with the 1-$\sigma$ error on the average
    inferred from the standard deviation of the probability from the
    individual MCMCs.  As discussed in the text, the power-law and
    Gaussian models are the most favored.}
\end{figure}

We have performed a suite of 500 independent reversible-jump MCMCs
jumping between all the models (both parametric and non-parametric)
described in Section \ref{sec:models} using the single-model MCMC
samples to construct an efficient jump proposal for each model as
described above (see Appendix \ref{sec:reversible-jump-mcmc}).  The
numbers of counts in each model are consistent across the MCMCs in the
suite; Figure \ref{fig:rj} displays the average probability for each
model across the suite, along with the 1-$\sigma$ errors on the
average inferred from the standard deviation of the model counts
across the suite.  Table \ref{tab:rj} gives the numerical values of
the average probability for each model across the suite of MCMCs.

The most favored model is the power law from Section
\ref{sec:power-law}, followed by the Gaussian model from Section
\ref{sec:gaussian}.  Interestingly, the theoretical curve from
\citet{Fryer2001} (the exponential model of Section
\ref{sec:exponential}) places fourth in the ranking of evidence.

\begin{table}
  \begin{center}
    \begin{tabular}{|l|r|}
      \hline
      Model & Relative Evidence \\
      \hline \hline
      Power Law (Section \ref{sec:power-law}) & 0.331488 \\
      \hline
      Gaussian (Section \ref{sec:gaussian}) & 0.288129 \\ 
      \hline
      Log Normal (Section \ref{sec:log-normal}) & 0.138435 \\
      \hline
      Exponential (Section \ref{sec:exponential}) & 0.0916218 \\
      \hline
      Two Gaussian (Section \ref{sec:gaussian}) & 0.0662577 \\
      \hline
      Histogram (1 Bin, Section \ref{sec:non-parametric-models}) &
      0.0641941 \\
      \hline
      Histogram (2 Bin, Section \ref{sec:non-parametric-models}) &
      0.015184 \\
      \hline 
      Histogram (3 Bin, Section \ref{sec:non-parametric-models}) &
      0.00332933 \\
      \hline
      Histogram (4 Bin, Section \ref{sec:non-parametric-models}) &
      0.000999976 \\
      \hline 
        Histogram (5 Bin, Section \ref{sec:non-parametric-models}) &
      0.0003614  \\
      \hline      
    \end{tabular}
  \end{center}
  \caption{\label{tab:rj} Relative probabilities of the various models
    from Section \ref{sec:models} implied by the low-mass data.  (See also Figure \ref{fig:rj}.)  These probabilities have been 
    computed from reversible-jump
    MCMC samples using the efficient jump proposal algorithm in Appendix \ref{sec:reversible-jump-mcmc}.}
\end{table}

Though the model probabilities presented in this section have small
statistical error, they are subject to large ``systematic'' error.
The source of this error is both the particular choice of model prior
(uniform across models) and the choice of priors on the parameters
within each model used for this work.  For example, the
theoretically-preferred exponential model (Section
\ref{sec:exponential}) is only a factor of $\sim 3$ away from the
power law model (Section \ref{sec:power-law}), which does not have
theoretical support.  Is such support worth a factor of three in the
model prior?  Alternately, we may say we know (in advance of any mass
measurements) that black holes must exist with mass $\lesssim
10\Msun$; then we could, for example, impose a prior on the minimum
mass in the exponential model ($\Mmin$) that is uniform between $0$
and $10 \Msun$, which would reduce the prior volume available for the
model by a factor of 4 without significantly reducing the posterior
support for the model.  This has the same effect as increasing the
model prior by a factor of 4, which would move this model from fourth
to first place.  Of course, we would then have to modify the prior
support for the other models to take into account the restriction that
there must be black holes with $M \lesssim 10\Msun$....
\citet{Linder2008} discuss these issues in the context of cosmological
model selection, concluding with a warning against over-reliance on
model selection probabilities.

Nevertheless, we believe that our model comparison is reasonably fair
(see the discussion of priors in Section \ref{sec:priors}).  It seems
safe to conclude that ``single-peaked'' models (the power-law and
Gaussian) are preferred over ``extended'' models (the exponential or
log-normal), or those with ``structure'' (the many-bin histograms or
two-Gaussian model).  Previous studies have also supported the
``single, narrow peak'' mass distribution \citep{Bailyn1998,Ozel2010}.
In this light, poor performance of the single-bin histogram is
surprising.

\subsection{Combined Sample}
\label{sec:high-mass}

\begin{figure}
  \begin{center}
    \plotone{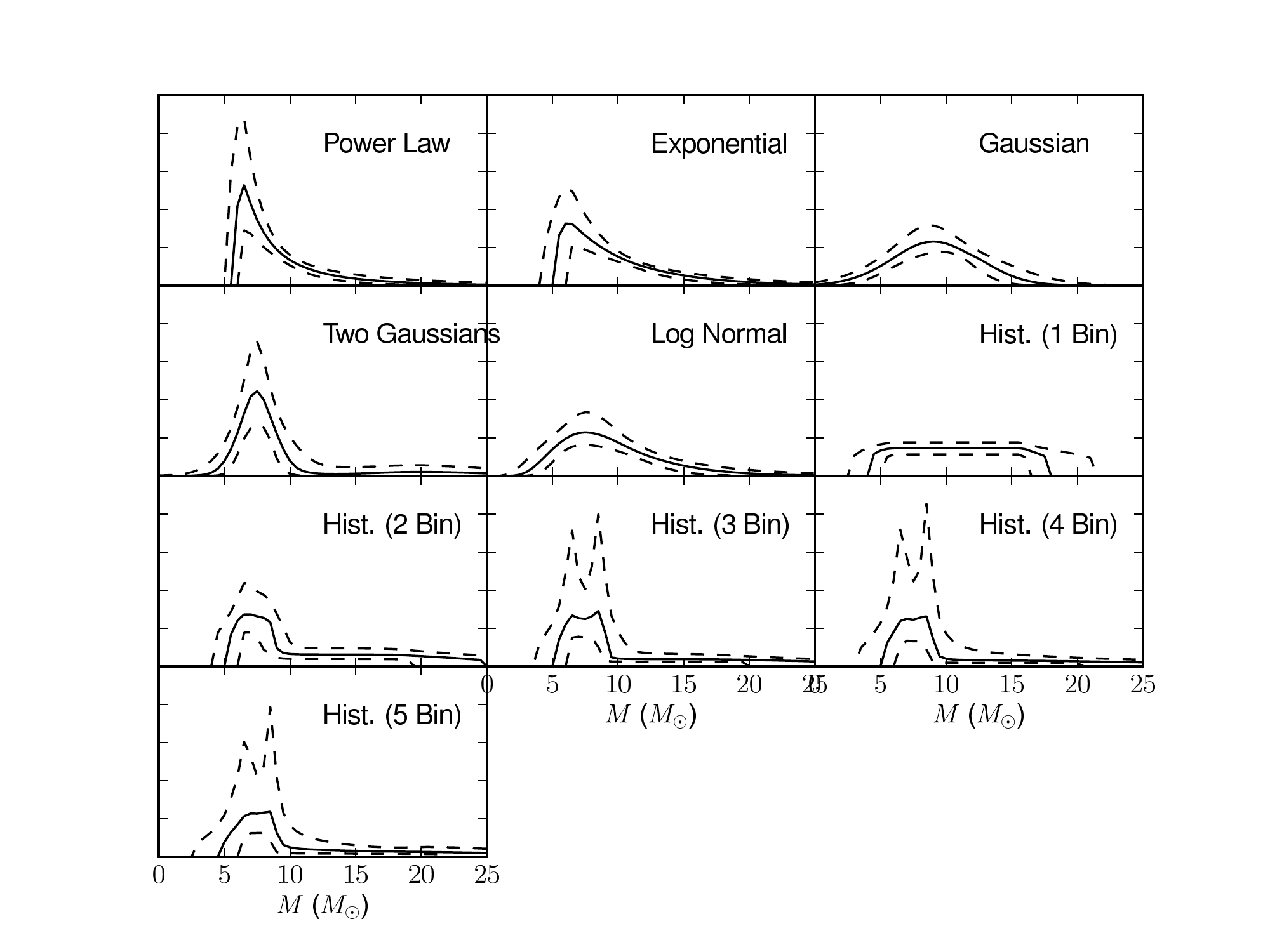}
  \end{center}
  \caption{\label{fig:high-mass-dists} The median (solid line), 10\%
    (lower dashed line), and 90\% (upper dashed line) values of the
    black hole mass distribution, $p(M|\theta)$, at various masses
    implied by the posterior $p(\theta|d)$ for the models discussed in
    Sections \ref{sec:parametric-models} and
    \ref{sec:non-parametric-models}.  These distributions use the
    combined sample of 20 observations in Table \ref{tab:sources},
    including the high-mass, wind-fed systems.  Note that these
    ``distributions of distributions'' are not necessarily normalized,
    and need not be ``shaped'' like the underlying model
    distributions.  Compare to Figure \ref{fig:dists}, which includes
    only the low-mass systems in the analysis.  Including the
    high-mass systems tends to widen the distribution toward the
    high-mass end and, in models that allow it, produce a second,
    high-mass peak in addition to the one in Figure
    \ref{fig:dists}. }
\end{figure}

This section repeats the analysis of the models from Section
\ref{sec:models}, but including the high-mass, wind-fed systems from
Table \ref{tab:sources} (see also Figure \ref{fig:high-masses}) in the
sample.  Figure \ref{fig:high-mass-dists} displays bounds on the value
of the underlying mass distribution for the various models in Section
\ref{sec:models} applied to this data set; compare to Figure
\ref{fig:dists}.  The inclusion of the high-mass, wind-fed systems
tends to widen the distribution toward the high-mass end and, in
models that allow it, produce a second, high-mass peak in addition
to the one in Figure \ref{fig:dists}.

\subsubsection{Power Law}

\begin{figure}
  \begin{center}
    \plotone{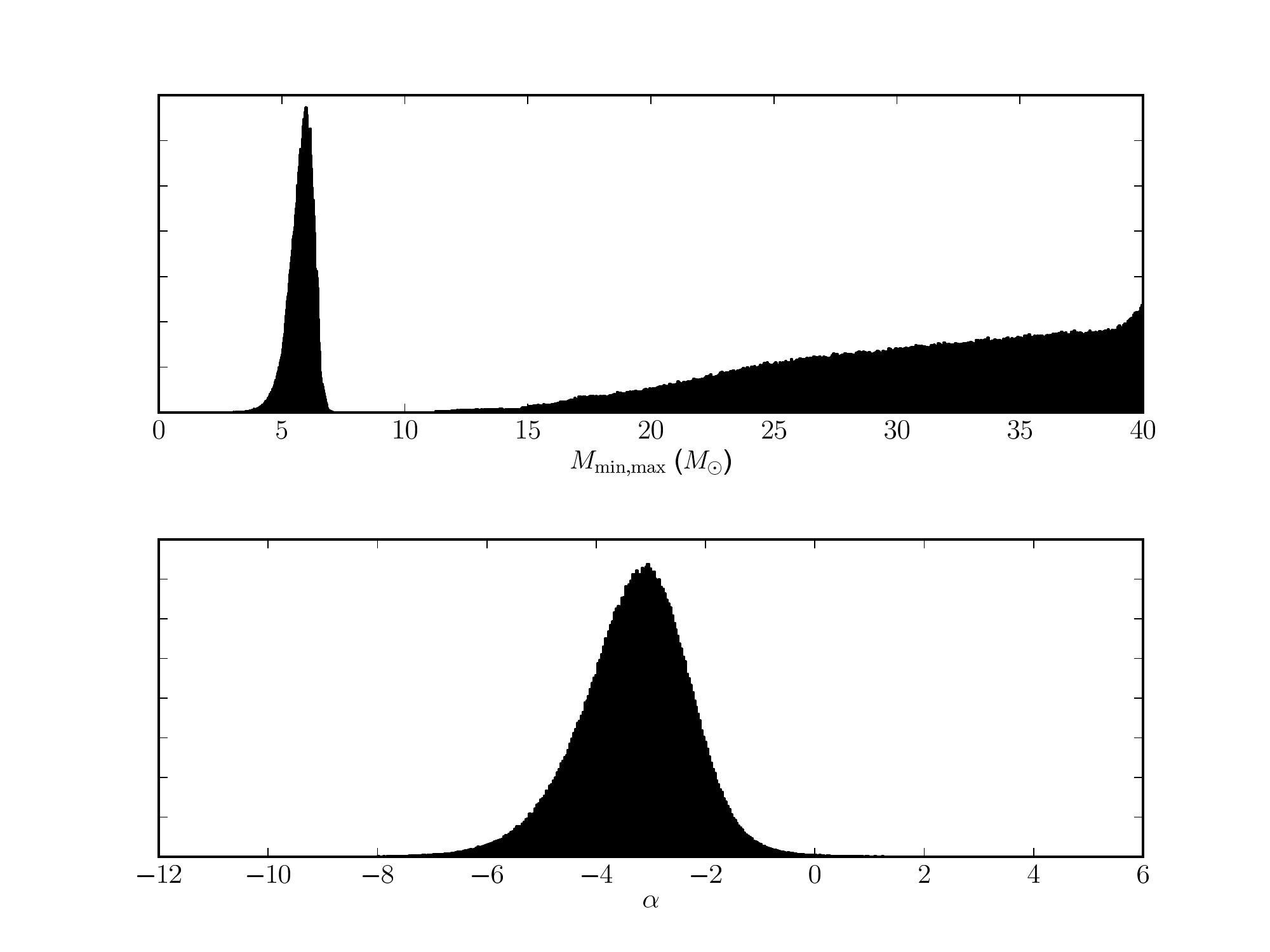}
  \end{center}
  \caption{\label{fig:power-law-high} Histograms of the marginalized
    distribution for the three parameters $\Mmin$ (top, left), $\Mmax$
    (top, right), and $\alpha$ (bottom) from the power-law model
    including the high-mass samples in the MCMC.  The distribution for
    $\Mmax$ is quite broad because the best fit power laws slope
    downward ($\alpha < 0$), making this parameter less relevant.  The
    range $-5.05 \leq \alpha \leq -1.77$ encloses 90\% of the
    probability; the median value of $\alpha$ is -3.23.  The presence
    of the high-mass samples in the analysis produces a distinctive
    tail, eliminating the correlations discussed in Section
    \ref{sec:power-law} and displayed in Figure \ref{fig:power-law-2D}
    for the low-mass subset of the observations. }
\end{figure}

Figure \ref{fig:power-law-high} presents the marginalized
distribution for the three power-law parameters $\Mmin$, $\Mmax$, and
$\alpha$ (Section \ref{sec:power-law}) from an analysis including the
high-mass systems.  The distribution for $\Mmax$ is quite broad
because the best fit power laws slope downward ($\alpha < 0$), making
this parameter less relevant.  The range $-5.05 \leq \alpha \leq
-1.77$ encloses 90\% of the probability; the median value of $\alpha$
is -3.23.  The presence of the high-mass samples in the analysis
produces a distinctive tail, eliminating the correlations discussed in
Section \ref{sec:power-law} and displayed in Figure
\ref{fig:power-law-2D} for the low-mass subset of the observations.

\subsubsection{Decaying Exponential}

\begin{figure}
  \begin{center}
    \plotone{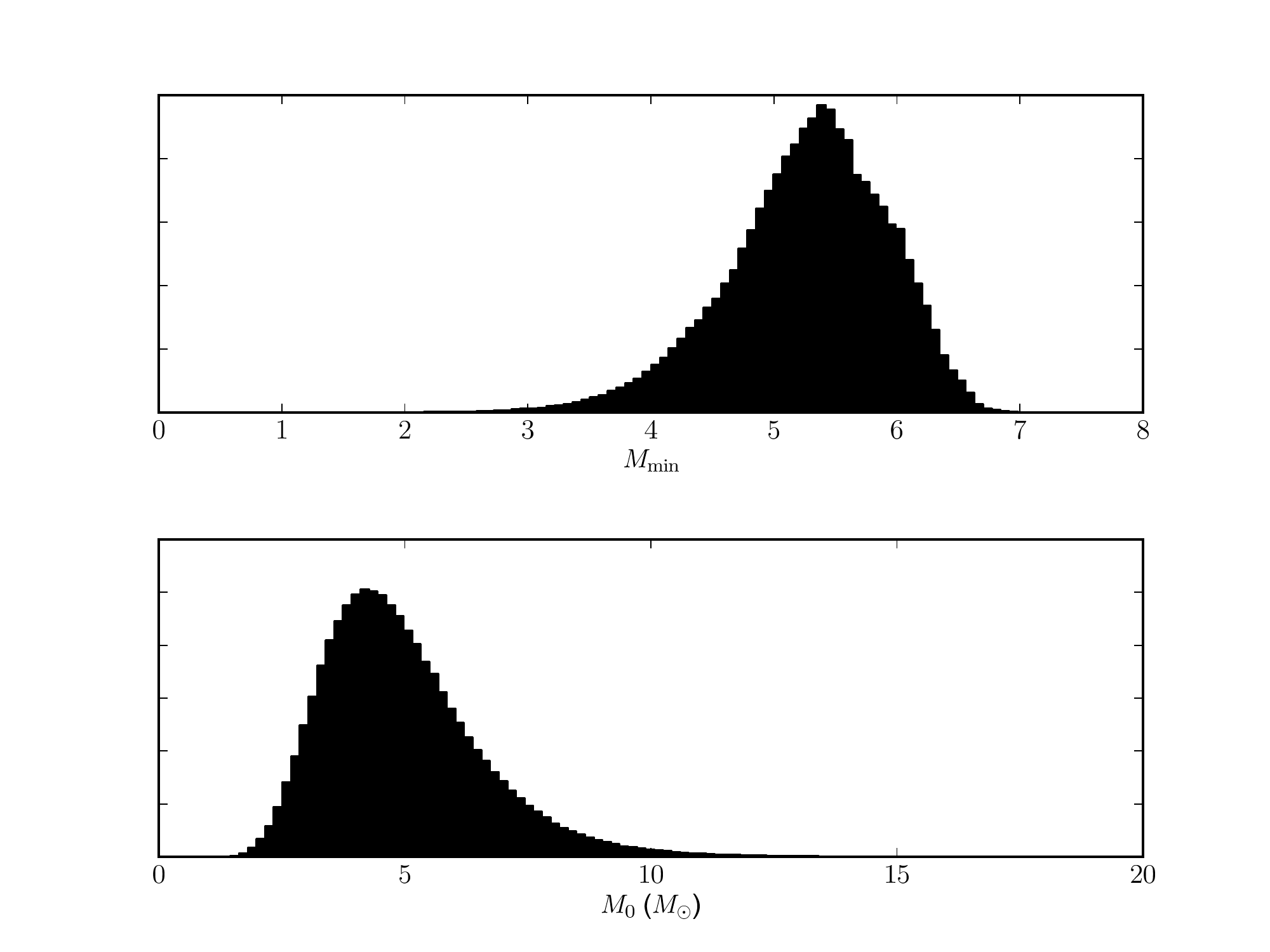}
  \end{center}
  \caption{\label{fig:exp-cutoff-high} The marginalized distributions
    for the exponential parameters $\Mmin$ (top) and $M_0$ (bottom)
    defined in Section \ref{sec:exponential} from an analysis
    including the high-mass systems.  The distribution for the scale
    mass, $M_0$, has moved to higher masses relative to Figure
    \ref{fig:exp-marginal} to fit the tail of the mass distribution;
    we now have $2.8292 \leq M_0 \leq 7.9298$ with 90\% confidence,
    with median 4.7003.  The distribution for $\Mmin$ is less
    affected, though it has broadened somewhat toward low masses.}
\end{figure}

Figure \ref{fig:exp-cutoff-high} displays the marginalized
distributions for the exponential parameters $\Mmin$ and $M_0$
(Section \ref{sec:exponential}) from an analysis including the
high-mass systems.  The distribution for the scale mass, $M_0$, has
moved to higher masses relative to Figure \ref{fig:exp-marginal} to
fit the tail of the mass distribution; the distribution for $\Mmin$ is
less affected, though it has broadened somewhat toward low masses.

\subsubsection{Gaussian}

\begin{figure}
  \begin{center}
    \plotone{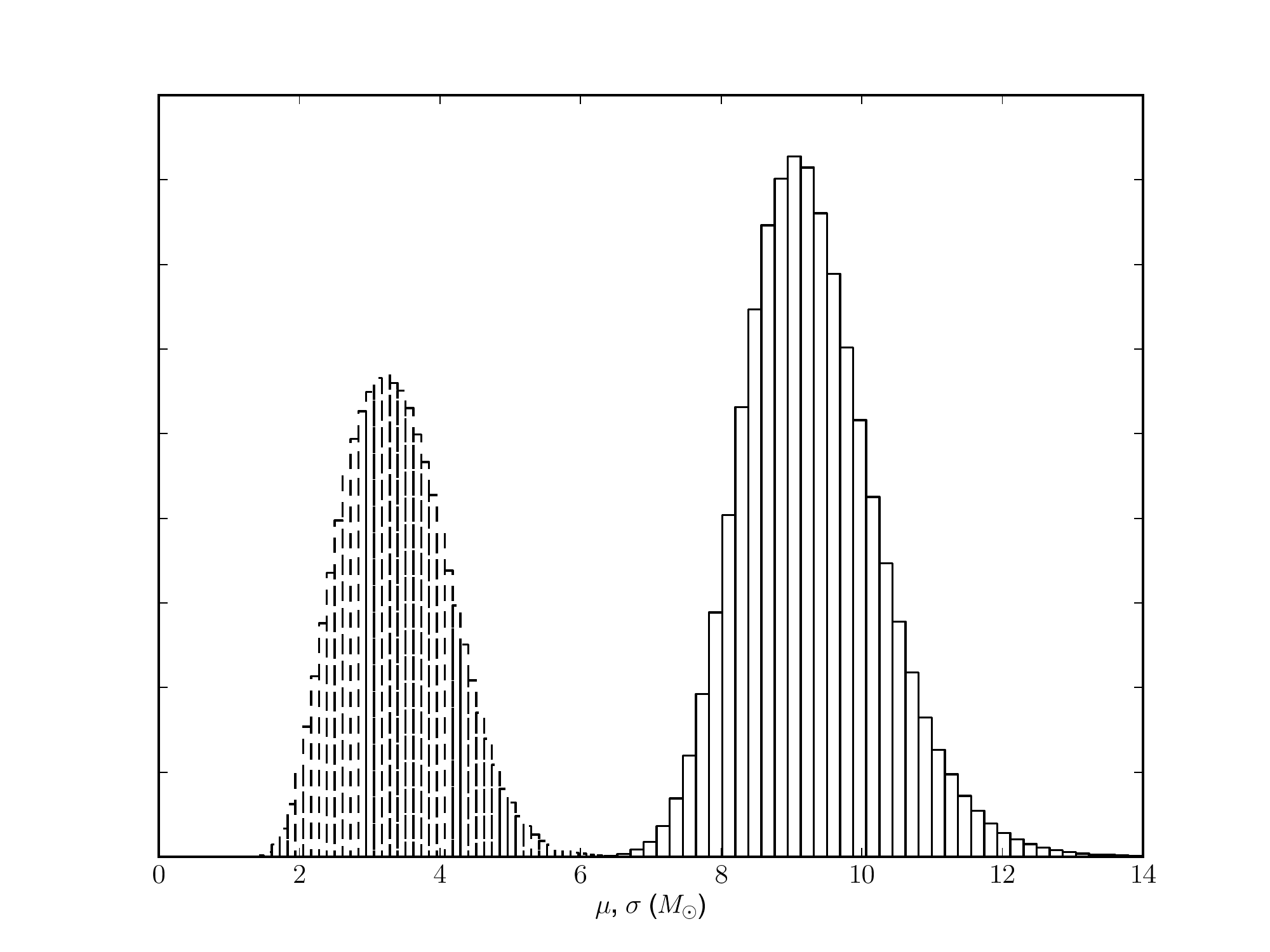}
  \end{center}
  \caption{\label{fig:gaussian-high} The marginalized distributions
    for the Gaussian parameters when the high-mass objects are
    included in the mass distribution.  The mean mass, $\mu$ (solid
    histogram), and the mass standard deviation, $\sigma$ (dashed
    histogram), are both increased relative to Figure
    \ref{fig:gaussian} to account for the broader distribution and
    high-mass tail.  The peak of the underlying mass distribution
    lies in the range $7.8660 \leq \mu \leq 10.9836$ with 90\%
    confidence; the median value is 9.2012.}
\end{figure}

Figure \ref{fig:gaussian-high} displays the marginalized distributions
for the Gaussian parameters (Section \ref{sec:gaussian}) when the
high-mass objects are included in the mass distribution.  The mean
mass, $\mu$, and the mass standard deviation, $\sigma$, are both
increased relative to Figure \ref{fig:gaussian} to account for the
broader distribution and high-mass tail.

\subsubsection{Two Gaussian}

\begin{figure}
  \begin{center}
    \plotone{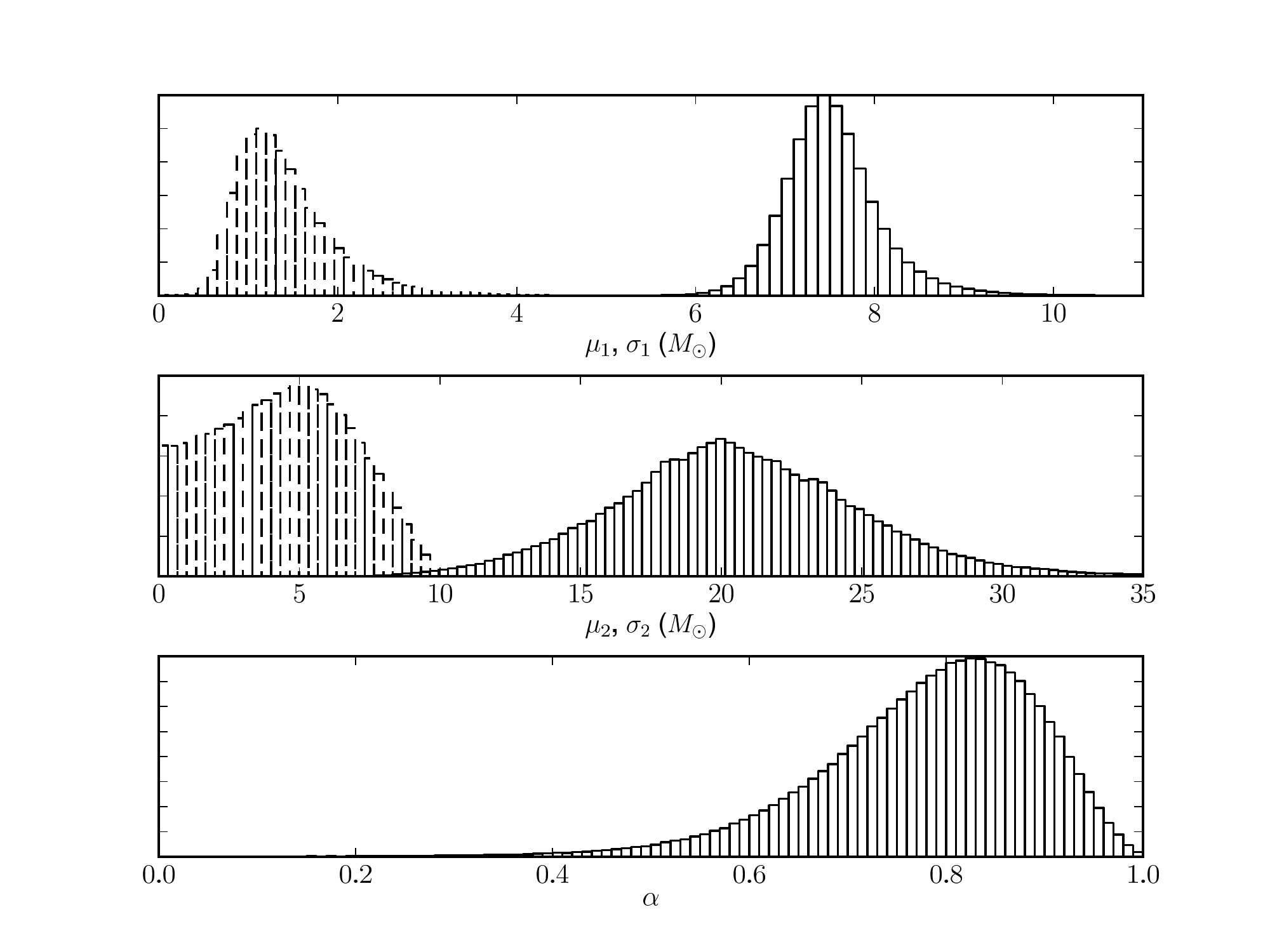}
  \end{center}
  \caption{\label{fig:two-gaussian-high} The marginalized
    distributions for the two-Gaussian parameters (Section
    \ref{sec:gaussian}) when the high-mass samples are included in the
    analysis.  The means ($\mu_1$ and $\mu_2$) are represented by the
    solid histograms; the standard deviations ($\sigma_1$ and
    $\sigma_2$) are represented by the dashed histograms.  In stark
    contrast to Figure \ref{fig:two-gaussian}, there are two
    well-defined, separated peaks; the low-mass peak reproduces the
    results from the low-mass samples, while the high-mass peak
    ($13.5534 \leq \mu_2 \leq 27.9481$ with 90\% confidence; median
    20.3839) matches the new high-mass samples.  The peak in $\alpha$
    near 0.8 is consistent with approximately 15 out of 20 samples
    belonging to the low-mass peak.}
\end{figure}

The analysis of the two-Gaussian model shows the largest change when
the high-mass samples are included.  Figure
\ref{fig:two-gaussian-high} shows the marginalized distributions for
the two-Gaussian parameters (Section \ref{sec:gaussian}) when the
high-mass samples are included in the analysis.  In stark contrast
to Figure \ref{fig:two-gaussian}, there are two well-defined,
separated peaks; the low-mass peak reproduces the results from the
low-mass samples, while the high-mass peak ($13.5534 \leq \mu_2 \leq
27.9481$ with 90\% confidence; median 20.3839) matches the new
high-mass samples.  The peak in $\alpha$ near 0.8 is consistent with
approximately 4/5 the total probability being concentrated in the 15
low-mass samples.

\subsubsection{Log Normal}

\begin{figure}
  \begin{center}
    \plotone{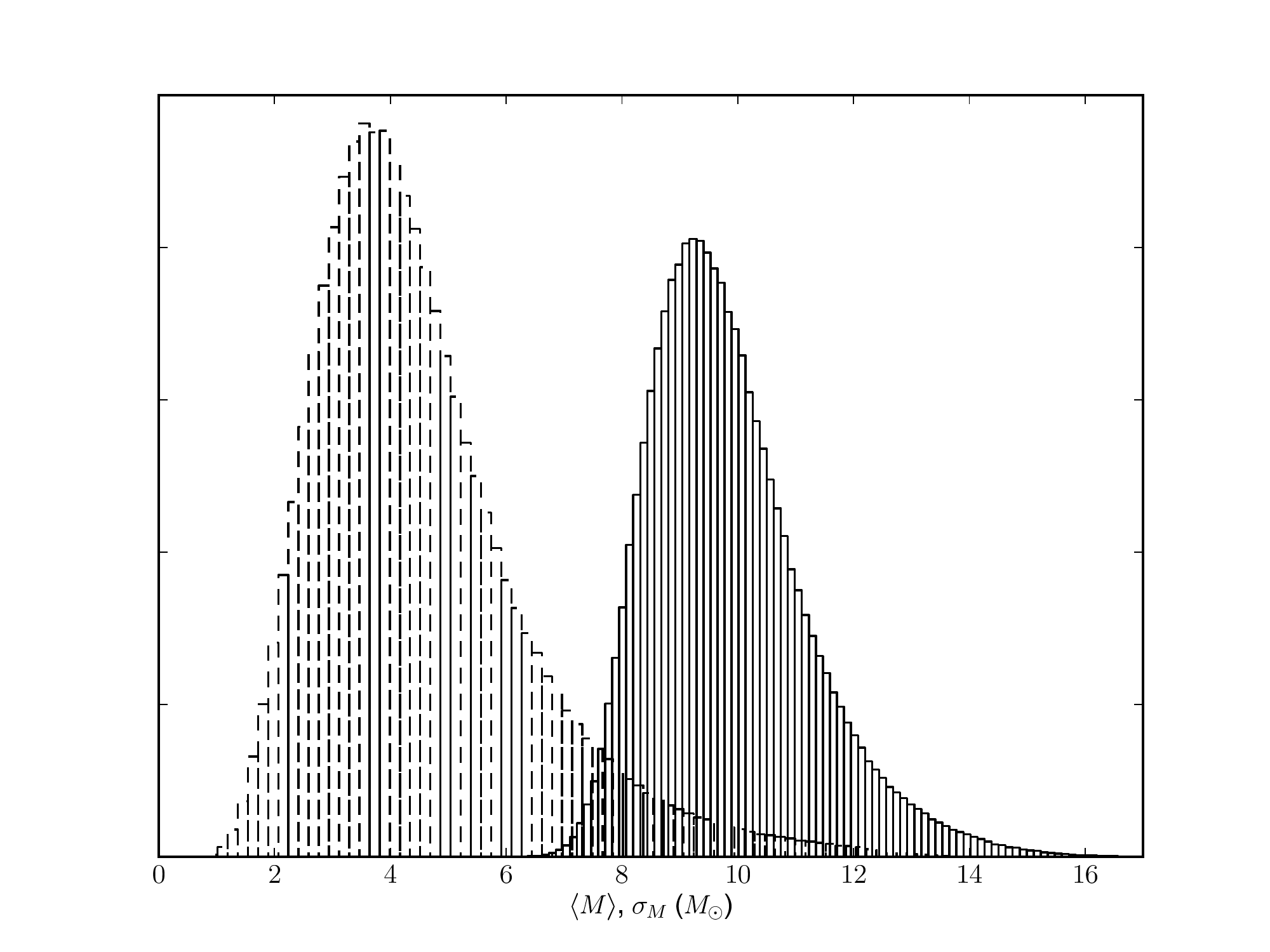}
  \end{center}
  \caption{\label{fig:log-normal-high} The marginalized distributions
    for the log-normal parameters (Section \ref{sec:log-normal};
    $\langle M \rangle$ solid, $\sigma_M$ dashed) when the high-mass
    samples are included in the analysis.  The changes when the
    high-mass samples are included (compare to Figure
    \ref{fig:log-normal}) are similar to the changes in the Gaussian
    distribution: the mean mass moves to higher masses, and the
    distribution broadens.}
\end{figure}

The marginalized distributions for the log-normal parameters (Section
\ref{sec:log-normal}) when the high-mass samples are included in the
analysis are displayed in Figure \ref{fig:log-normal-high}.  The
changes when the high-mass samples are included (compare to Figure
\ref{fig:log-normal}) are similar to the changes in the Gaussian
distribution: the mean mass moves to higher masses, and the
distribution broadens.  Because the log-normal distribution is
inherently asymmetric, with a high-mass tail, it does not need to
widen as much as the Gaussian distribution did.

The confidence limits on the parameters for the parametric models of
the underlying mass distribution are displayed in Table
\ref{tab:high-mass-parametric} (compare to Table
\ref{tab:low-mass-parametric}).

\begin{table}
  \begin{center}
    \begin{tabular}{|l|c|c|c|c|c|c|}
      \hline
      Model & Parameter & 5\% & 15\% & 50\% & 85\% & 95\% \\
      \hline \hline
      Power Law (Equation \eqref{eq:power-law-dist}) & $\Mmin$ & 
      4.87141 & 5.29031 & 5.85019 & 6.26118 & 6.45674 \\
      \hline
      & $\Mmax$ & 19.1097 & 23.4242 & 31.5726 & 37.7519 & 39.3369 \\
      \hline
      & $\alpha$ & -5.04879 & -4.30368 & -3.23404 & -2.31365 & -1.77137 \\
      \hline \hline
      Exponential (Equation \eqref{eq:exp-def})& $\Mmin$ & 
      4.0865 & 4.60236 & 5.32683 & 5.94097 & 6.22952 \\
      \hline
       & $M_0$ & 2.82924 & 3.41139 & 4.70034 & 6.52214 & 7.92979 \\
      \hline \hline
      Gaussian (Equation \eqref{eq:gaussian-def}) & $\mu$ & 
      7.86599 & 8.33118 & 9.20116 & 10.2493 & 10.9836 \\
      \hline
      & $\sigma$ & 2.23643 & 2.58899 & 3.33545 & 4.17886 & 4.67881 \\
      \hline \hline
      Two Gaussian (Equation \eqref{eq:two-gaussian-def}) & $\mu_1$ & 
      6.741 & 7.02724 & 7.48174 & 8.0139 & 8.46626 \\
      \hline
      & $\mu_2$ & 13.5534 & 16.202 & 20.3839 & 24.9259 & 27.9481 \\
      \hline
      & $\sigma_1$ & 0.742824 & 0.913941 & 1.31244 & 1.94862 & 2.50238 \\
      \hline
      & $\sigma_2$ & 0.511159 & 1.5025 & 4.39824 & 7.04612 & 8.25905 \\
      \hline
      & $\alpha$ & 0.575692 & 0.670978 & 0.798227 & 0.891522 & 0.932143 \\
      \hline \hline
      Log Normal (Equation \eqref{eq:log-normal-def}) & $\langle M \rangle$ & 
      8.00086 & 8.51192 & 9.6264 & 11.1851 & 12.3986 \\
      \hline
      & $\sigma_M$ & 2.19262 & 2.8137 & 4.16742 & 6.25101 & 8.11839 \\
      \hline
    \end{tabular}
  \end{center}
  \caption{\label{tab:high-mass-parametric} Quantiles of the
    marginalized distribution for each of the parameters in the models
    discussed in Section \ref{sec:parametric-models} when the
    high-mass samples are included in the analysis (compare to Table
    \ref{tab:low-mass-parametric}).  We indicate the 5\%, 15\%, 50\% 
    (median), 85\%, and 95\% quantiles.}
\end{table}

\subsubsection{Histogram Models}

The non-parametric (histogram; see Section
\ref{sec:non-parametric-models}) models also show evidence of a long
tail from the inclusion of the high-mass samples.  Table
\ref{tab:high-mass-non-parametric} displays confidence limits on the
histogram parameters for the analysis including the high-mass
systems; compare to Table \ref{tab:low-mass-non-parametric}.

\begin{table}
  \begin{center}
    \begin{tabular}{|c|c|c|c|c|c|c|}
      \hline
      Bins & Boundary & 5\% & 15\% & 50\% & 85\% & 95\% \\
      \hline \hline
      \hline
      1 & $w_0$ & 2.22294 & 3.12695 & 4.2456 & 5.15132 & 5.58265 \\
      \hline
      & $w_1$ & 15.93 & 16.2535 & 17.7836 & 20.5449 & 22.5836 \\
      \hline \hline
      2 & $w_0$ & 3.87202 & 4.49983 & 5.41234 & 6.08334 & 6.35933 \\
      \hline
      & $w_1$ & 7.22163 & 8.25079 & 8.93669 & 9.71551 & 10.4287 \\
      \hline
      & $w_2$ & 18.4762 & 19.9798 & 24.941 & 32.5972 & 36.8615 \\
      \hline \hline
      3 & $w_0$ & 3.39289 & 4.24509 & 5.41694 & 6.15087 & 6.42822 \\
      \hline
      & $w_1$ & 6.41849 & 6.71984 & 7.47263 & 8.2942 & 8.61785 \\
      \hline
      & $w_2$ & 8.41449 & 8.64664 & 9.17056 & 10.4075 & 12.2718 \\
      \hline
      & $w_3$ & 18.5705 & 21.0481 & 27.1494 & 34.7753 & 38.0652 \\
      \hline \hline
      4 & $w_0$ & 2.42094 & 3.69875 & 5.2596 & 6.25449 & 6.54316 \\
      \hline
      & $w_1$ & 5.83725 & 6.2836 & 6.84987 & 7.8033 & 8.27706 \\
      \hline
      & $w_2$ & 6.94919 & 7.43628 & 8.38531 & 9.13401 & 9.91845 \\
      \hline
      & $w_3$ & 8.50371 & 8.75188 & 9.86694 & 17.1848 & 22.1086 \\
      \hline
      & $w_4$ & 18.5823 & 21.4628 & 28.367 & 35.8118 & 38.5278 \\      
      \hline \hline
      5 & $w_0$ & 1.73691 & 3.19184 & 4.89769 & 5.9547 & 6.35522 \\
      \hline
      & $w_1$ & 5.46124 & 5.95881 & 6.59431 & 7.26795 & 7.91821 \\
      \hline
      & $w_2$ & 6.63468 & 6.9804 & 7.93239 & 8.60918 & 9.06926 \\
      \hline
      & $w_3$ & 7.89654 & 8.35634 & 8.91766 & 10.6568 & 13.9644 \\
      \hline
      & $w_4$ & 8.74064 & 9.42672 & 15.8004 & 22.7101 & 27.6399 \\
      \hline
      & $w_5$ & 20.0202 & 22.9065 & 29.6307 & 36.6606 & 38.8573 \\
      \hline
    \end{tabular}
  \end{center}
  \caption{\label{tab:high-mass-non-parametric} The 5\%, 15\%, 50\%
    (median), 85\%, and 95\% quantiles for the bin boundaries in the
    one- through five-bin histogram models discussed in Section
    \ref{sec:non-parametric-models} in an 
    analysis including the high-mass, wind-fed systems.  
    The tails evident in Figure \ref{fig:high-mass-dists} are apparent
    here as well; compare to Table \ref{tab:low-mass-non-parametric}.}
\end{table}

\subsubsection{Model Selection for the Combined Sample}
\label{sec:high-mass-model-selection}

Repeating the model selection analysis discussed in Section
\ref{sec:low-mass-model-selection} for the sample including the
high-mass systems, we find that the model probabilities have changed
with the inclusion of the extra five systems.  As before, we assume
for this analysis that the model priors are equal.

\begin{figure}
  \begin{center}
    \plotone{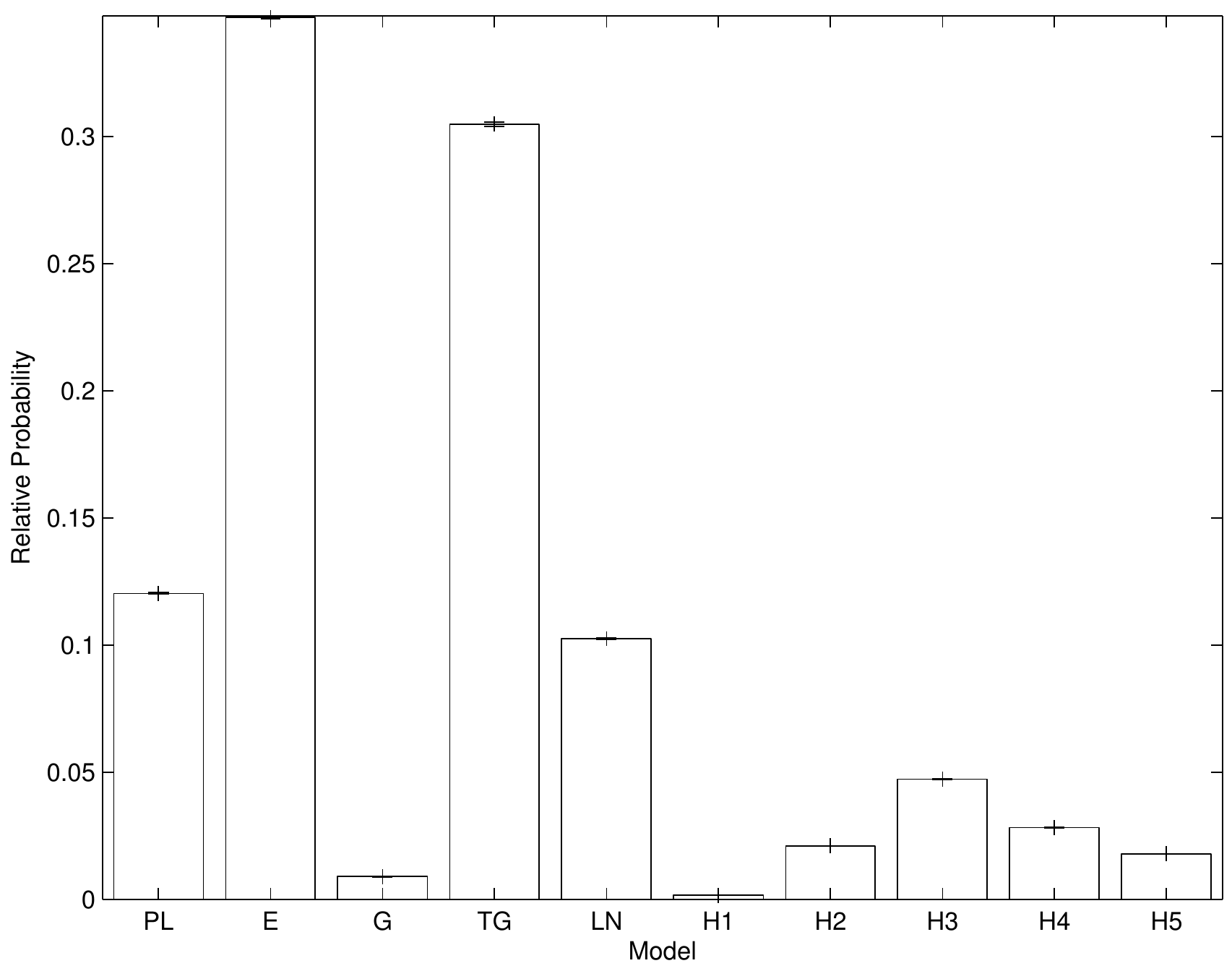}
  \end{center}
  \caption{\label{fig:high-rj-evidence} The relative probability of
    the models discussed in Section \ref{sec:models} as computed using
    the reversible-jump MCMC with the efficient jump proposal
    algorithm described in Section \ref{sec:reversible-jump-mcmc},
    applied to all 20 systems in Table \ref{tab:sources} (i.e.\
    including the high-mass systems).  (See also Table
    \ref{tab:rj-high}.)  In increasing order along the $x$-axis, the
    models are the power-law of Section \ref{sec:power-law} (PL), the
    decaying exponential of Section \ref{sec:exponential} (E), the
    single Gaussian of Section \ref{sec:gaussian} (G), the double
    Gaussian of Section \ref{sec:gaussian} (TG), and the one-, two-,
    three-, four-, and five-bin histogram models of Section
    \ref{sec:non-parametric-models} (H1, H2, H3, H4, H5,
    respectively).  The average of 500 independent reversible-jump
    MCMCs is plotted, along with the 1-$\sigma$ error on the average
    inferred from the standard deviation of the probability from the
    individual MCMCs.  Compare to Figure \ref{fig:rj}.}
\end{figure}

Reversible jump MCMC calculations of the model probabilities are
displayed in Figure \ref{fig:high-rj-evidence}; compare Figure
\ref{fig:rj}.  The relative model probabilities are given in Table
\ref{tab:rj-high}.  The exponential model is the most favored model
for the combined sample, with the two-Gaussian model the second-most
favored.  The ranking of models differs significantly from the
low-mass samples.  The improvement of the exponential model relative
to the low-mass analysis is encouraging for theoretical calculations
that attempt to model the entire population of X-ray binaries with
this mass model.  Note also that the increasing structure of the mass
distribution favors histogram models with three bins over those with
fewer bins.

\begin{table}
  \begin{center}
    \begin{tabular}{|l|r|}
      \hline
      Model & Relative Evidence \\
      \hline \hline
      Exponential (Section \ref{sec:exponential}) & 0.346944 \\
      \hline
      Two Gaussian (Section \ref{sec:gaussian}) & 0.304923 \\
      \hline
      Power Law (Section \ref{sec:power-law}) & 0.120313 \\
      \hline
      Log Normal (Section \ref{sec:log-normal}) & 0.102536 \\
      \hline
      Histogram (3 Bin, Section \ref{sec:non-parametric-models}) &
      0.0473464 \\
      \hline
      Histogram (4 Bin, Section \ref{sec:non-parametric-models}) &
      0.0282086 \\
      \hline 
      Histogram (2 Bin, Section \ref{sec:non-parametric-models}) &
      0.0210994 \\
      \hline 
      Histogram (5 Bin, Section \ref{sec:non-parametric-models}) &
      0.0179703  \\
      \hline      
      Gaussian (Section \ref{sec:gaussian}) & 0.00901719 \\ 
      \hline
      Histogram (1 Bin, Section \ref{sec:non-parametric-models}) &
      0.00164214 \\
      \hline
    \end{tabular}
  \end{center}
  \caption{\label{tab:rj-high} Relative probabilities of the various models
    from Section \ref{sec:models} implied by the combined sample of systems.  (See also Figure \ref{fig:high-rj-evidence}.)  These probabilities have been 
    computed from reversible-jump MCMC samples using the efficient jump proposal algorithm in Appendix \ref{sec:reversible-jump-mcmc}.}
\end{table}

\section{The Minimum Black Hole Mass}
\label{sec:minimum-mass}

It is interesting to use our models for the underlying black hole mass
distribution in X-ray binaries to place constraints on the minimum
black hole mass implied by the present sample.  \citet{Bailyn1998}
addressed this question in the context of a ``mass gap'' between the
most massive neutron stars and the least massive black holes.  The
more recent study of \citet{Ozel2010} also looked for a mass gap using
a subset of the models and systems presented here.  Both works found
that the minimum black hole mass is significantly above the maximum
neutron star mass \citep{Kalogera1996} of $\sim 3 \Msun$ (though
\citet{Ozel2010} only state their evidence for a gap in terms of the
maximum-posterior parameters and not the full extent of their
distributions).

\begin{figure}
  \begin{center}
    \plotone{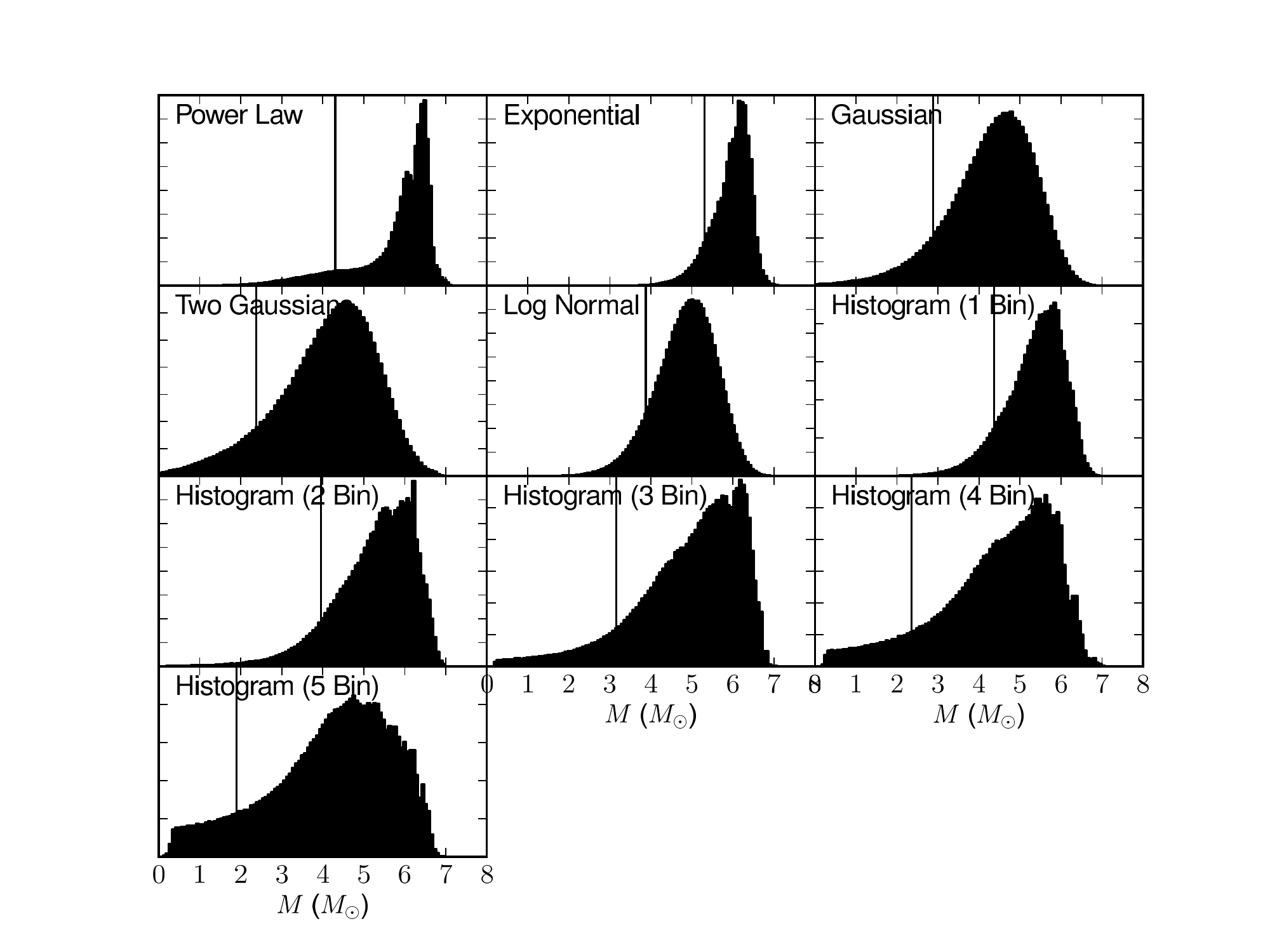}
  \end{center}
  \caption{\label{fig:min-mass} The distributions for the minimum
    black hole mass, $M_{1\%}$, calculated from the MCMC samples for
    the models in Section \ref{sec:models} applied to the low-mass
    systems.  For the most favored models, the power-law and Gaussian,
    the 90\% confidence limit on the minimum black hole mass is 4.3
    $\Msun$ and 2.9 $\Msun$, respectively.  In all plots, we indicate
    the 90\% confidence bound (i.e.\ the 10\% quantile) on the minimum
    black hole mass with a vertical line.}
\end{figure}

The distributions of the minimum black hole mass from the analysis of
the low-mass samples are displayed in Figure \ref{fig:min-mass}.  The
minimum black hole mass is defined as the 1\% mass quantile,
$M_{1\%}$, of the black-hole mass distribution (i.e.\ the mass lying
below 99\% of the mass distribution).  (A quantile-based definition is
necessary in the case of those distributions that do not have a hard
cutoff mass; even for those that do, like the power-law model, it can
be useful to define a ``soft'' cutoff in the event that the lower mass
hard cutoff becomes an irrelevant parameter as discussed in Section
\ref{sec:power-law}.)  For each mass distribution parameter sample
from our MCMC, we can calculate the distribution's minimum black hole
mass; the collection of these minimum black hole masses approximates
the distribution of minimum black hole masses implied by the data in
the context of that distribution.  Figure \ref{fig:min-mass} plots
histograms of the minimum black hole mass samples.

We find that the best-fit model for the low-mass systems (the
power-law) has $M_{1\%} > 4.3$ $\Msun$ in 90\% of the MCMC samples
(i.e.\ at 90\% confidence).  This is significantly above the maximum
theoretically-allowed neutron star mass, $\sim 3 \Msun$ (e.g.\
\citet{Kalogera1996}).  Hence we conclude that the low-mass systems
show strong evidence of a mass gap.

\begin{figure}
\begin{center}
    \plotone{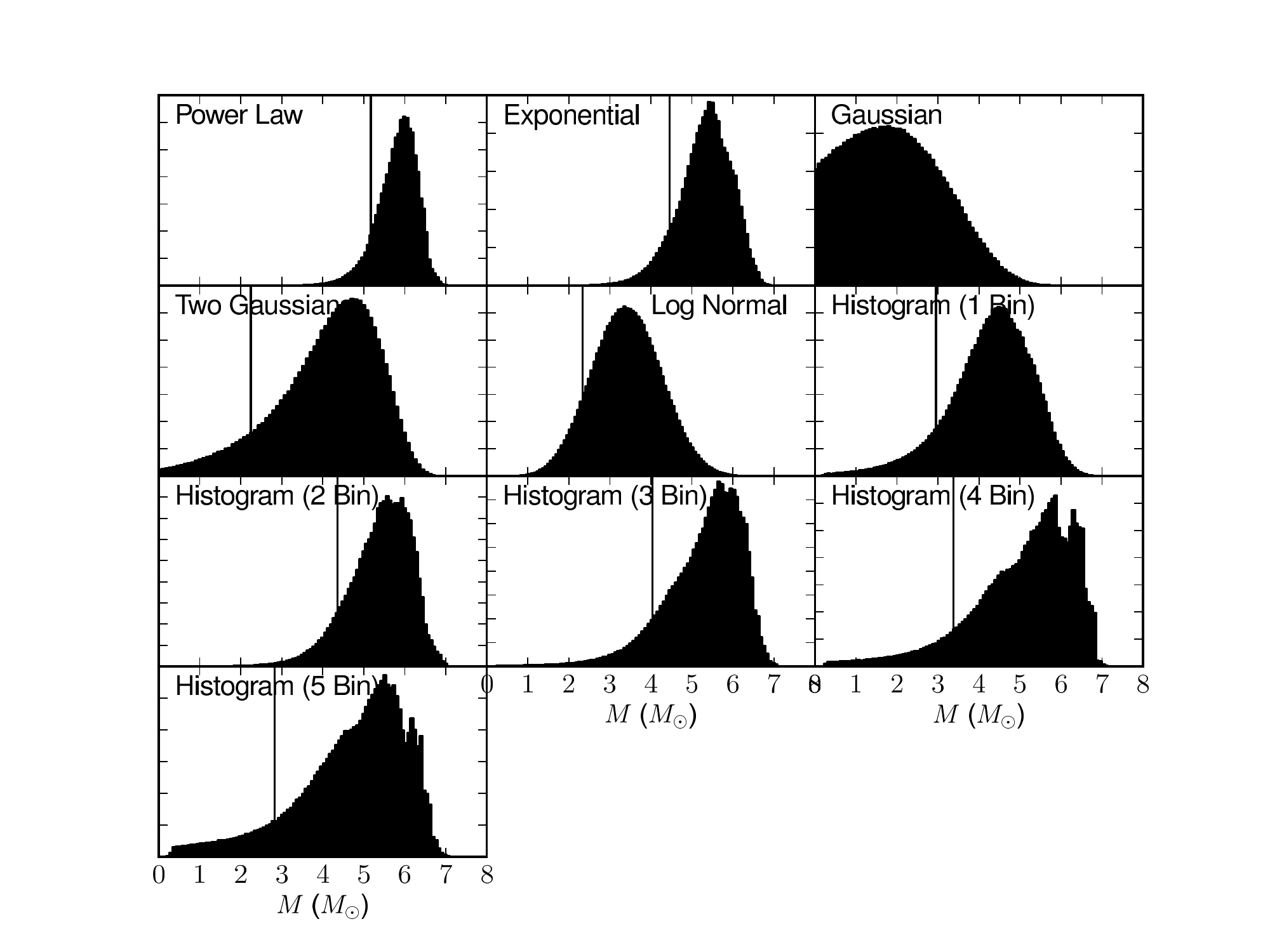}
  \end{center}
  \caption{\label{fig:high-min-mass} The distributions for the minimum
    black hole mass, $M_{1\%}$, calculated from the MCMC samples for
    the models in Section \ref{sec:models} using the combined sample
    of systems.  For the two most favored models, the exponential and
    two-Gaussian, the 90\% confidence limit on the minimum black hole
    mass is 4.5 $\Msun$ and 2.3 $\Msun$, respectively.  For every
    model, we indicate the 90\% confidence bound on the minimum black
    hole mass with a vertical line.}
\end{figure}

The distribution of minimum black hole masses for the analysis of the
combined sample (i.e.\ including the high-mass systems) is shown in
Figure \ref{fig:high-min-mass}.  For the most favored model, the
exponential, we find that $M_{1\%} > 4.5$ $\Msun$ with 90\%
confidence.  We therefore conclude that there is strong evidence for a
mass gap in the combined sample as well.

Table \ref{tab:mmin-quants} gives the 10\%, 50\% (median), and 90\%
quantiles for the minimum black hole mass implied by the low-mass
sample; Table \ref{tab:mmin-quants-high} gives the same, but for the
combined sample of systems.

\begin{table}
  \begin{center}
    \begin{tabular}{|l|l|c|r|}
      \hline
      Model & 10\% & 50\% & 90\% \\
      \hline \hline
      Power Law (Section \ref{sec:power-law}) & 4.3 & 6.1 & 6.6 \\
      \hline
      Gaussian (Section \ref{sec:gaussian}) & 2.9 & 4.4 & 5.5 \\
      \hline 
      Log Normal (Section \ref{sec:log-normal}) & 3.9 & 4.9 & 5.8 \\
      \hline 
      Exponential (Section \ref{sec:exponential}) & 5.3 & 6.0 & 6.5 \\
      \hline 
      Two Gaussian (Section \ref{sec:gaussian}) & 2.4 & 4.2 & 5.5 \\
      \hline 
      Histogram (1 Bin, Section \ref{sec:non-parametric-models}) & 4.4 & 5.5 & 6.2 \\
      \hline
      Histogram (2 Bin, Section \ref{sec:non-parametric-models}) & 4.0 & 5.4 & 6.3 \\
      \hline 
      Histogram (3 Bin, Section \ref{sec:non-parametric-models}) & 3.2 & 5.2 & 6.3 \\
      \hline
      Histogram (4 Bin, Section \ref{sec:non-parametric-models}) & 2.4 & 4.7 & 6.0 \\
      \hline
      Histogram (5 Bin, Section \ref{sec:non-parametric-models}) & 1.9 & 4.4 & 6.0 \\
      \hline
    \end{tabular}
  \end{center}
  \caption{\label{tab:mmin-quants} The 10\%, 50\% (median), and 90\% quantiles for the minimum black hole mass (in units of $\Msun$) implied by the low-mass sample in the context of the various models for the black hole mass distribution.  The models are listed in order of preference from model selection (Section \ref{sec:low-mass-model-selection}, Figure \ref{fig:rj}, and Table \ref{tab:rj}).}
\end{table}

\begin{table}
  \begin{center}
    \begin{tabular}{|l|l|c|r|}
      \hline
      Model & 10\% & 50\% & 90\% \\
      \hline \hline
      Exponential (Section \ref{sec:exponential}) & 4.5 & 5.4 & 6.1 \\
      \hline
      Two Gaussian (Section \ref{sec:gaussian}) & 2.3 & 4.3 & 5.5 \\
      \hline
      Power Law (Section \ref{sec:power-law}) & 5.1 & 5.9 & 6.4 \\
      \hline
      Histogram (3 Bin, Section \ref{sec:non-parametric-models}) & 4.0 & 5.5 & 6.3 \\
      \hline 
      Histogram (4 Bin, Section \ref{sec:non-parametric-models}) & 3.4 & 5.3 & 6.4 \\
      \hline
      Histogram (2 Bin, Section \ref{sec:non-parametric-models}) & 4.4 & 5.5 & 6.2 \\
      \hline
      Histogram (5 Bin, Section \ref{sec:non-parametric-models}) & 2.8 & 5.0 & 6.2 \\
      \hline
      Gaussian (Section \ref{sec:gaussian}) & -0.64 & 1.4 & 3.4 \\
      \hline 
      Histogram (1 Bin, Section \ref{sec:non-parametric-models}) & 2.9 & 4.4 & 5.5 \\
      \hline
    \end{tabular}
  \end{center}
  \caption{\label{tab:mmin-quants-high} The 10\%, 50\% (median), and 90\% quantiles for the distribution of minimum black hole masses (in units of $\Msun$) implied by the combined sample in the context of the various models for the black hole mass distribution.  The models are listed in order of preference from model selection (Section \ref{sec:high-mass-model-selection}, Figure \ref{fig:high-rj-evidence}, and Table \ref{tab:rj-high}). }
\end{table}

\section{Summary and Conclusion}
\label{sec:conclusion}

We have presented a Bayesian analysis of the mass distribution of
stellar-mass black holes in X-ray binary systems.  We considered
separately a sample of 15 low-mass, Roche lobe-filling systems and a
sample of 20 systems containing the 15 low-mass systems and five
high-mass, wind-fed X-ray binaries.  We used MCMC methods to sample
the posterior distributions of the parameters implied by the data for
five parametric models and five non-parametric (histogram) models for
the mass distribution.  For both sets of samples, we used reversible
jump MCMCs (exploiting a new algorithm for efficient jump proposals in
such calculations) to perform model selection on the suite of models.
The consideration of a broad range of models and the model-selection
analysis, along with consideration of the full posterior distribution
on the minimum black hole mass, significantly expand earlier
statistical analyses of black hole mass measurements
\citep{Bailyn1998,Ozel2010}.

For the low-mass systems, we found the limits on model parameters in
Tables \ref{tab:low-mass-parametric} and
\ref{tab:low-mass-non-parametric}.  The relative model probabilities
from the model selection are given in Table \ref{tab:rj}.  The most
favored model for the low-mass systems is a power law.  The equivalent
limits on the model parameters for the combined systems are given in
Tables \ref{tab:high-mass-parametric} and
\ref{tab:high-mass-non-parametric}.  Unlike the low-mass systems, the
most favored model for the combined sample is the exponential model.
This difference indicates that the low-mass subsample is not
consistent with being drawn from the distribution of the combined
population.

We found strong evidence for a mass gap between the most massive
neutron stars and the least massive black holes.  For the low-mass
systems, the most favored, power law model gives a black hole mass
distribution whose 1\% quantile lies above $4.3 \Msun$ with 90\%
confidence.  For the combined sample of systems, the most favored,
exponential model gives a black hole mass distribution whose 1\%
quantile lise above $4.5 \Msun$ with 90\% confidence.  Although the
study methodology was different, the existence of a mass gap was
pointed out first by \citet{Bailyn1998} and most recently by
\citet{Ozel2010} (who did not consider a power law model, and applied
both Gaussian and exponential models to the low-mass systems, where
the exponential is strongly disfavored compared to our power-law
model).

Theoretical expectations for the black hole mass distribution have
been examined in \citet{Fryer2001}.  They considered results of
supernova explosion and fallback simulations \citep{Fryer1999} applied
to single star populations; they also included a heuristic treatment
of the possible effects of binary evolution on the black hole mass
distribution.  It is interesting that we find the most-favored model
for the combined sample to be an exponential, as discussed by
\citet{Fryer2001}.  On the other hand, we find the most-favored model
for the low-mass sample to be a power law, with the exponential model
strongly disfavored for this sample. In agreement with
\citet{Bailyn1998} and \citet{Ozel2010}, we too conclude that both the
low-mass and combined samples require the presence of a gap between 3
and 4--4.5 $\Msun$.

\citet{Fryer1999} discussed two possible causes of such a gap: (1) a
step-like dependence of supernova energy on progenitor mass or (2)
selection biases.  Current simulations of core collapse in massive
stars may shed light on the dependence of supernova energy on
progenitor mass.  Selection biases can occur because the X-ray
binaries with very low-mass black holes systems are more likely to be
persistently Roche-lobe overflowing, preventing dynamical mass
measurements.  \citet{Ozel2010} conclude that the presence of such
biases is not enough to account for the gap, arguing that the number
(26) of observed persistent X-ray sources not known to be neutron
stars is insufficient to populate the 2--5 $\Msun$ region of any black
hole mass distribution that rises toward low masses.  Population
synthesis models incorporating sophisticated treatment of binary
evolution and transient behavior (e.g.\ \citet{Fragos2008,Fragos2009})
could help shed light on this possibility.

\acknowledgements

WMF, NS, and VK are supported by NSF grants CAREER AST-0449558 and
AST–0908930.  AC, LK, and CB are supported by NSF grant
NSF/AST-0707627.  IM acknowledges support from the NSF AAPF under
award AST-0901985.  Calculations for this work were performed on the
Northwestern Fugu cluster, which was partially funded by NSF MRI grant
PHY-0619274.  We thank Jonathan Gair for helpful discussions.

\appendix

\section{Markov Chain Monte Carlo}
\label{sec:mcmc}

MCMC methods produce a Markov chain (or sequence) of parameter
samples, $\{ \vtheta_i \, | \, i = 1, \ldots \}$, such that a particular
parameter set, $\vtheta$, appears in the sequence with a frequency
equal to its probability according to a posterior, $p(\vtheta|d)$.  A
Markov chain has the property that the transition probability from one
element to the next, $p(\vtheta_i \to \vtheta_{i+1})$, depends only on
the value of $\vtheta_i$, not on any previous values in the chain.

One way to produce a sequence of MCMC samples is via the following
algorithm, first proposed by \citet{Metropolis1953} and used widely in
the physical sciences thereafter:
\begin{enumerate}
  \item Begin with the current sample, $\vtheta_i$.
  \item Propose a new sample, $\vtheta_p$, by drawing randomly from a
    ``jump proposal distribution'' with probability $Q(\vtheta_i \to
    \vtheta_p)$.  Note that $Q(\vtheta_i \to \vtheta_p)$ can depend on
    the current parameters, $\vtheta_i$, and any other ``constant''
    data, but cannot examine the history of the chain beyond the most
    recent point.  This is necessary to preserve the Markovian
    property of the chain.
  \item Compute the ``acceptance'' probability,
    \begin{equation}
      \label{eq:paccept}
      p_{\textnormal{accept}} \equiv
      \frac{p(\vtheta_p|d)}{p(\vtheta_i|d)} \frac{Q(\vtheta_p \to
        \vtheta_i)}{Q(\vtheta_i \to \vtheta_p)}
    \end{equation}
  \item With probability $\min(1,p_{\textnormal{accept}})$ ``accept''
    the proposed $\vtheta_p$, setting $\vtheta_{i+1} = \vtheta_p$;
    otherwise set $\vtheta_{i+1} = \vtheta_i$.
\end{enumerate}
This algorithm is more likely to accept a proposed jump when it
increases the posterior (the first factor in Equation
\eqref{eq:paccept}) and when it is to a location in parameter space
from which it is easy to return (the second factor in Equation
\eqref{eq:paccept}); the combination of these influences in Equation
\eqref{eq:paccept} ensures that the equilibrium distribution of the
chain is $p(\vtheta|d)$.  As $i \to \infty$ the samples $\vtheta_i$
are distributed according to $p(\vtheta|d)$.  

In practice the number of samples required before the chain
appropriately samples $p(\vtheta|d)$ depends strongly on the jump
proposal distribution; proposal distributions that often propose jumps
toward or within regions of large $p(\vtheta|d)$ can be very
efficient, while poor proposal distributions can require prohibitively
large numbers of samples before convergence.  

There is no foolproof test for the convergence of a chain.  In this
work, we test the convergence of our chains several ways.  The most
basic is by comparing the statistics calculated from the entire chain
to statistics calculated from only the first half of the chain; when
the chain has converged, the two calculations agree.  This is a
necessary, but not sufficient, condition for convergence.

\begin{figure}
  \begin{center}
    \plotone{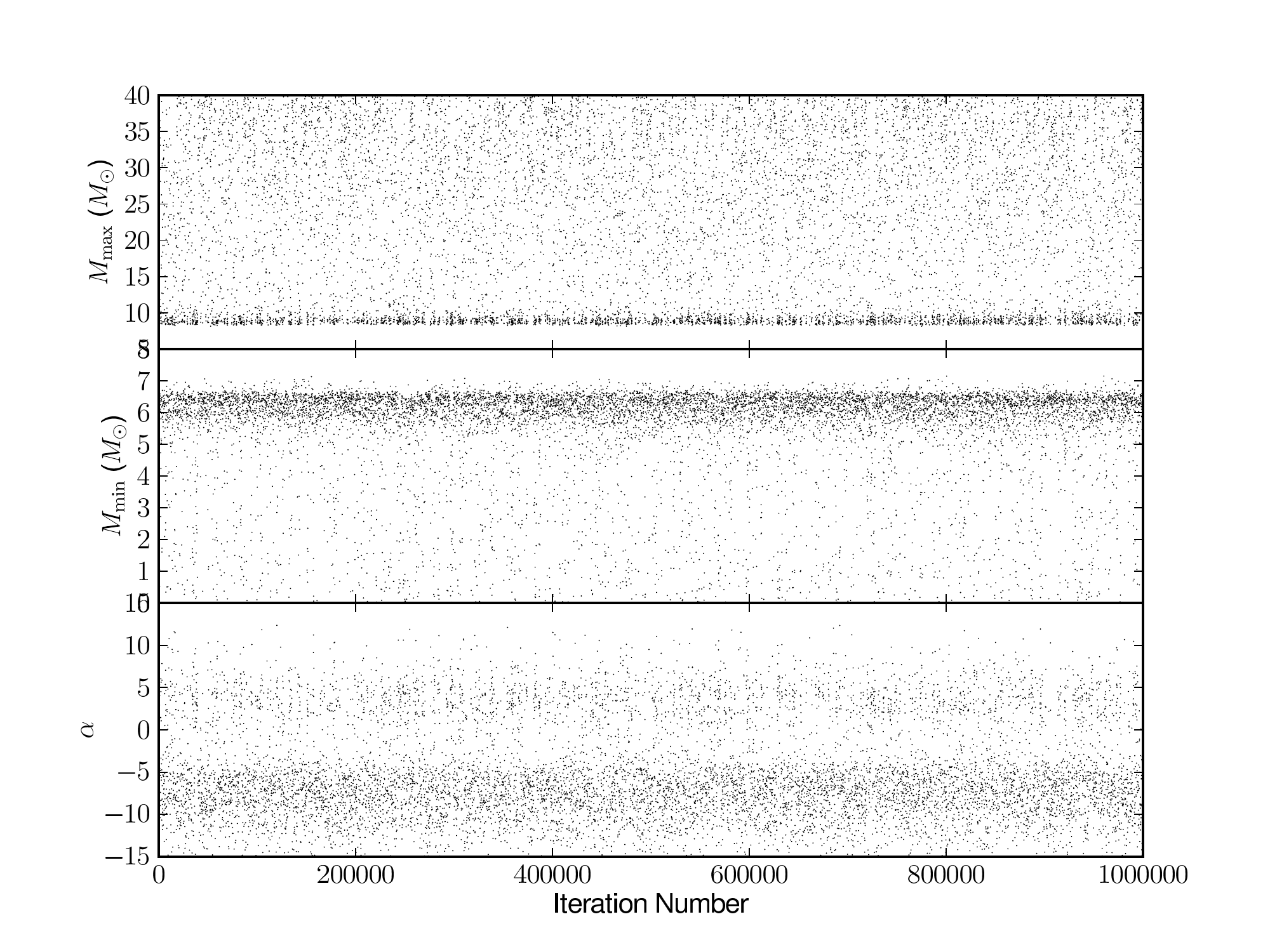}
  \end{center}
  \caption{\label{fig:sample-trace} Parameter sample traces from the
    MCMC applying the power-law model to the low-mass systems (the
    parameter histograms for this MCMC appear in Figure
    \ref{fig:power-law}).  For clarity, only every 100th sample point
    is plotted.  The chain is well-converged---it samples the regions
    of posterior support densely and randomly, without any visible
    trends or sticking points.}
\end{figure}

We also have examined the sample traces from our chains, to see that
the chains have densely and randomly sampled parameter space.  A
representative sample trace from our MCMC for the power-law model
applied to the low-mass systems appears in Figure
\ref{fig:sample-trace}.  Sample traces from MCMCs with other models
are similar.

Finally, for our most quantitative test of convergence we use the {\tt
  gibbsit} code to implement the Raftery-Lewis convergence test for
our quantile measures \citep{Raftery1992,Raftery1992a,Raftery1995}.
The most extreme quantile is the most difficult to determine
accurately because---by design---there will be fewer samples in the
tail than in the main body of a distribution obtained from an MCMC.
Accordingly, we focus on the 90\% confidence limit on the minimum
black hole mass.  For a quantile, $q$, the Raftery-Lewis test attempts
to estimate how many samples from an MCMC are needed to determine $q$
to within $\pm r$ at a confidence $s$.  We use $r = 0.0125$ and $s =
0.95$.  The Raftery-Lewis test approximates the MCMC chain as a
two-state Markov chain, the two states being ``within the quantile in
question'' and ``outside the quantile in question.''  The 2x2
transition matrix for this two-state Markov chain and the associated
uncertainty can be calculated analytically
\citep{Raftery1992,Raftery1992a,Raftery1995}, allowing the algorithm
to determine the number of sample points required before the quantile
of interest is determined sufficiently accurately.  

For our chains, in the worst case (the power-law on the lower-mass
samples, as shown in Figures \ref{fig:power-law} and
\ref{fig:sample-trace}), we have twice as many samples as the
Raftery-Lewis convergence test estimates we need to determine the 90\%
confidence level on the minimum mass; for all the other chains, we
have about 20 times as many samples as the Raftery-Lewis criterion
estimates are required.  We suspect that the slow convergence of the
power-law model on the lower-mass systems is due to the long tails in
the mass parameters and the width of the distribution on the power-law
exponent.  In any case, the Raftery-Lewis test indicates that all our
chains are converged sufficiently to determine the 90\% quantile to
within $\sim 1\%$.

We begin the chain at an arbitrary point in parameter space; this is
equivalent to taking a finite section of an infinite chain that begins
with the chosen point.  Every point in parameter space occurs in an
infinite chain, and no section of the chain is better than any other,
so a sufficiently long, but finite, section of the infinite chain
chosen in this manner can be representative of the statistics of the
chain as a whole.  However, because consecutive samples in a chain are
correlated with each other, the beginning of our finite chain has a
``memory'' of the starting point; we discard enough points at the
beginning of the finite chain that we can be confident it does not
retain a memory of the arbitrary starting point.  The points discarded
in this way are commonly called ``burn-in'' points.

\section{Reversible-Jump MCMC}
\label{sec:reversible-jump-mcmc}

The algorithm described here for the reversible jump MCMC we have used
for the model comparison in this work is more fully described in
\citet{Farr2010}.  In particular, \citet{Farr2010} demonstrates the
efficiency gains from the algorithm, and shows experimentally that the
algorithm indeed provides a consistent RJMCMC, with the correct
relative model probabilities for the case where the Bayes factor
between models can be calculated analytically.

Consider the problem of model selection among a set of models, and the
``super-model'' that encompasses all the models under consideration.
The parameter space of the super-model consists of a discrete
parameter that identifies the choice of model, $M_i$, and the
continuous parameters appropriate for this model, $\vtheta_i$.  We
denote a point in the super-model parameter space by $\{M_i,
\vtheta_i\}$; each such point is a statement that, e.g., ``the
underlying mass distribution for black holes in the galaxy is a
Gaussian, with parameters $\mu$ and $\sigma$,'' or ``the underlying
mass distribution for black holes in the galaxy is a triple-bin
histogram with parameters $w_1$, $w_2$, $w_3$, and $w_4$,'' or ....
To compare models, we are interested in the quantity (see Equation
\eqref{eq:model-posterior-def})
\begin{equation}
  p(M_i|d) = \int d\vtheta_i p(M_i, \vtheta_i|d).
\end{equation}
If we perform an MCMC in the super-model parameter space, then we
obtain a chain of samples $\{M_i, \vtheta_i \, | \, i = 1, \ldots\}$
distributed in parameter space with density $p(M_i,\vtheta_i|d)
d\vtheta_i$ and we can estimate the integral as
\begin{equation}
    p(M_i|d) = \int d\vtheta_i p(M_i, \vtheta_i|d) \approx \frac{N_i}{N},
\end{equation}
where $N_i$ is the number of samples in the chain lying in the
parameter space of model $M_i$ and $N$ is the total number of samples
in the chain.  The fraction of samples lying in the parameter space of
model $M_i$ gives the probability of that model relative to the other
models under consideration.

To perform the MCMC in the super-model parameter space, we must
propose jumps not only between points in a particular model's
parameter space, but also between the parameter spaces of different
models.  For this MCMC to be efficient, proposed jumps into a model
from another should favor regions with large posterior; when the
posterior is highly-peaked in a small region of parameter space,
proposed jumps outside this region are unlikely to be accepted, and
the reversible-jump MCMC samples will require a very long chain to
properly sample the ``super-model'' posterior.

We can exploit the information we have from single-model MCMCs to
generate efficient jump proposal distributions for our reversible jump
MCMC.  We would like to propose jumps that roughly follow the
distribution of samples in the single-model MCMCs.  We can do this by
assigning a neighborhood to each point in the sample using an
algorithm we will describe in the following paragraphs; the
neighborhoods are non-overlapping, completely cover the region of
parameter space with prior support, and contain only one point from
the MCMC samples.  To propose a jump into model $M_i$, we choose a
point uniformly from that single-model MCMC and then propose a jump
drawn uniformly from that point's neighborhood.  This is equivalent to
drawing from a piecewise-constant approximation to the single-model
posterior, where each neighborhood contributes a constant fraction,
$1/N_i$, to the cumulative jump probability.  In regions of high
density the neighborhoods are smaller, and the jump probability
density is correspondingly higher.  Because the neighborhoods cover
the entire region of prior support, it is possible for the proposal to
propose any point in parameter space with prior support (though points
in regions of low single-model posterior are of course unlikely to be
proposed).

\begin{figure}
  \begin{center}
    \plotone{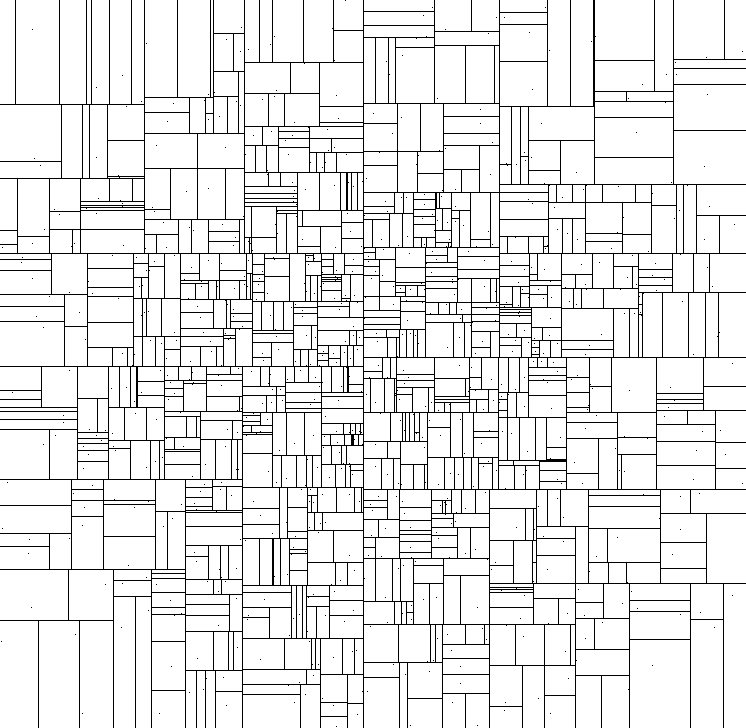}
  \end{center}
  \caption{\label{fig:kD-tree} The neighborhoods constructed from a
    two-dimensional kD-tree built from a sample of points with a
    Gaussian density distribution.  Each line on the figure
    corresponds to a sub-dividing box boundary drawn between the
    median of a subset of the sample points and its nearest
    neighbor. The peak of the Gaussian lies in the center of the
    figure; here the point density is highest and the neighborhoods
    are smallest.  Near the edges the density is lower, and the
    neighborhoods correspondingly larger.  The tree adapts itself to
    the local density of points.  If these were single-model MCMC
    samples, the corresponding jump proposal would first select one of
    the boxes uniformly at random, and then choose a point uniformly
    within the box to propose.  Since there are many more boxes near
    the center (each box corresponds to one point), and these boxes
    are smaller, the proposal will tend to concentrate its points
    there, approximately tracking the distribution of single-model
    MCMC samples.}
\end{figure}

To assign a neighborhood to each point in a single-model MCMC we use a
data structure called a kD-tree.  A kD-tree is a binary
space-partitioning tree.  To construct a kD-tree, we begin with the
set of points from a single-model MCMC and a box in parameter space
bounding the region of prior support (which must necessarily enclose
all the points).  The construction proceeds recursively: we choose a
dimension%
\footnote{Our algorithm chooses the dimension along which the
  numerical extent of the points is largest.  Other choices are
  possible; some algorithms cycle through the dimensions in order,
  while others choose a random dimension for each subdivision.  Our
  goal by picking the longest dimension is to produce neighborhoods
  that are ``square,'' at least in the chosen parametrization.} %
along which to divide the points, find the median point along that
dimension and its nearest neighbor, and divide the box at the midpoint
between these two points, producing two sub-boxes.  We then partition
the points into those to the left (i.e.\ smaller coordinate along the
given dimension) and right of the dividing line, and repeat this
procedure for each subset and the corresponding bounding box, until we
have only one point in each box.  An example of the neighborhoods that
result from a two-dimensional kD-tree constructed around a Gaussian
point distribution appears in Figure \ref{fig:kD-tree}.

Construction of a kD-tree is an $\order{N\log N}$ operation, where $N$
is the number of points in the tree.  Median finding is $\order{n}$,
where $n$ is the number of points from which the median is to be
obtained.  At level $i$ in the tree, there are $2^i$ subsets of
points, each of length $\order{N/2^i}$, so the total cost of the $2^i$
median calculations is $\order{N}$ at each level.  There are
$\order{\log N}$ levels in the tree, yielding a total construction
cost for the tree of $\order{N \log N}$.  

To find the neighborhood of a point using the tree, we begin at the
root of the tree, and examine the two sub-boxes at the next level
down.  The point will be in one of them; following that branch, we
have again two sub-boxes, one of which contains the point; following
that branch....  Eventually, the search terminates at a leaf of the
tree, containing the point in question.  The box at the leaf defines
the neighborhood of the point in the jump proposal algorithm described
above.  The total cost for this operation is proportional to the
number of levels in the tree, which is $\order{\log N}$.

In addition to the validation tests in \citet{Farr2010}, we have
validated our interpolation method with the following test.  We
imagine that we have a data set that can be fit by two models: an
``egg-crate'' model with likelihood
\begin{equation}
  L(\vtheta|d) = 2^N \prod_{i = 1}^N \sin^2 \left( 2\pi n \theta_i \right),
\end{equation}
and a single-Gaussian model with likelihood
\begin{equation}
  L(\vtheta|d) = \frac{1}{\left( 2\pi \right)^{N/2} \prod_{i=1}^N
    \sigma_i } \prod_{i = 1}^N \exp\left( - \frac{\left( \theta_i - \mu_i \right)^2}{2
      \sigma_i^2} \right),
\end{equation}
where the number of dimensions in each parameter space is $N = 5$, and
we choose $2n = 10$ peaks along each dimension for the egg-crate
model.  We restrict the parameter space to $\vtheta \in [0,1]^N$, and
choose a uniform prior on $\vtheta$ within that region.  We choose
$\mu_i = 1/2$, and $\sigma_i = 1/(20 i)$, so the Gaussian peak is
well-contained within the region of interest, being at least
$10\sigma$ away from the boundaries of the region.  From the point of
view of an (RJ)MCMC, the precise data that could produce such unusual
likelihoods from two models are irrelevant; the algorithm only cares
about the form of the likelihood and prior functions in parameter
space.  These likelihood functions provide a good test case for our
interpolation technique: the egg-crate model has a broad, multi-modal
likelihood, while the Gaussian model's likelihood is very concentrated
in a small region of parameter space.  Particularly for the Gaussian
model, an RJMCMC without interpolation---one that proposes inter-model
jumps from the prior, for example---would be extremely inefficient
because the region of parameter space with significant posterior
support is so small.

Our choice of likelihood and prior implies that the models have equal
evidence:
\begin{equation}
  p(d|M_i) = \int_{[0,1]^N} d^N \vtheta \, L(\vtheta|d) p(\vtheta) = 1.
\end{equation}
Using individual MCMC parameter samples to construct interpolations of
the single-model posterior, and running a $10^6$ sample
reversible-jump MCMC, we find
\begin{equation}
  \frac{p(d|\textnormal{egg crate})}{p(d|\textnormal{Gaussian})}
  \simeq \frac{N_{\textnormal{egg crate}}}{N_{\textnormal{Gaussian}}}
  = \frac{499285}{500715} \simeq 0.997,
\end{equation}
which has an error of $3\times 10^{-3}$, of order
$1/\sqrt{N_{\textnormal{egg crate}}} \sim
1/\sqrt{N_{\textnormal{Gaussian}}} \sim 1.4 \times 10^{-3}$, as would
be expected for 1M independent samples from a binomial distribution
with $p = 0.5$.  We conclude, as in \citet{Farr2010}, that our
interpolation method leads to accurate and efficient reversible-jump
MCMCs.

\bibliography{paper}

\end{document}